\documentclass[twocolumn]{aastex631}
\usepackage{mathtools, cuted}
\usepackage{amsmath}
\usepackage{lipsum, color}

\usepackage{multirow}
\usepackage{comment}

\shorttitle{Robust field-level inference with dark matter halos}
\shortauthors{Shao et al.}

\graphicspath{{./}{figures/}}

\begin{document}

\title{Robust field-level inference of cosmological parameters with dark matter halos}

\author[0000-0002-0152-6747]{Helen Shao}
\email{hshao@princeton.edu}
\affiliation{Department of Astrophysical Sciences, Princeton University, 4 Ivy Lane, Princeton, NJ 08544 USA}
\affiliation{Center for Computational Astrophysics, Flatiron Institute, 162 5th Avenue, New York, NY 10010, USA}

\author[0000-0002-4816-0455]{Francisco Villaescusa-Navarro}
\email{fvillaescusa@flatironinstitute.edu}
\affiliation{Center for Computational Astrophysics, Flatiron Institute, 162 5th Avenue, New York, NY 10010, USA}
\affiliation{Department of Astrophysical Sciences, Princeton University, 4 Ivy Lane, Princeton, NJ 08544 USA}

\author[0000-0002-0936-4279]{Pablo Villanueva-Domingo}
\affiliation{Computer Vision Center - Universitat Aut\`onoma de Barcelona (UAB), 08193 Bellaterra, Barcelona, Spain}

\author{Romain Teyssier}
\affiliation{Department of Astrophysical Sciences, Princeton University, 4 Ivy Lane, Princeton, NJ 08544 USA}

\author[0000-0002-9853-5673]{Lehman H. Garrison}
\affiliation{Center for Computational Astrophysics, Flatiron Institute, 162 5th Avenue, New York, NY 10010, USA}

\author{Marco Gatti}
\affiliation{Department of Physics and Astronomy, University of Pennsylvania, Philadelphia, PA 19104, USA}

\author[0000-0001-6039-9058]{Derek Inman}
\affiliation{Kavli Institute for the Physics and Mathematics of the Universe (WPI), The University of Tokyo Institutes \\for Advanced Study, The University of Tokyo, Kashiwa, Chiba 277-8583, Japan}

\author{Yueying Ni}
\affiliation{McWilliams Center for Cosmology, Department of Physics, Carnegie Mellon University, Pittsburgh, PA 15213, US}
\affiliation{Harvard-Smithsonian Center for Astrophysics, 60 Garden Street, Cambridge, MA 02138, US}

\author{Ulrich P. Steinwandel}
\affiliation{Center for Computational Astrophysics, Flatiron Institute, 162 5th Avenue, New York, NY 10010, USA}

\author[0000-0002-9789-6653]{Mihir Kulkarni}
\affiliation{Department of Physics and Astronomy and Ritter Astrophysical
Research Center, \\University of Toledo, 2801 W Bancroft Street, Toledo, OH 43606, USA}

\author[0000-0002-8365-0337]{Eli Visbal}
\affiliation{Department of Physics and Astronomy and Ritter Astrophysical
Research Center, \\University of Toledo, 2801 W Bancroft Street, Toledo, OH 43606, USA}

\author[0000-0003-2630-9228]{Greg L. Bryan}
\affiliation{Department of Astronomy, Columbia University, 550 West 120th Street, New York, NY, 10027, USA}

\author{Daniel Angl\'es-Alc\'azar}
\affiliation{Department of Physics, University of Connecticut, 196 Auditorium Road, U-3046, Storrs, CT, 06269, USA}
\affiliation{Center for Computational Astrophysics, Flatiron Institute, 162 5th Avenue, New York, NY 10010, USA}

\author[0000-0002-6292-3228]{Tiago Castro}
\affiliation{INAF-Osservatorio Astronomico di Trieste, Via G. B. Tiepolo 11, I-34131 Trieste, Italy}
\affiliation{INFN, Sezione di Trieste, Via Valerio 2, I-34127 Trieste TS, Italy}
\affiliation{IFPU, Institute for Fundamental Physics of the Universe, via Beirut 2, 34151 Trieste, Italy}

\author{Elena Hern\'andez-Mart\'inez}
\affiliation{Universit\"ats-Sternwarte, Fakult\"at f\"ur Physik, Ludwig-Maximilians-Universit\"at M\"unchen, Scheinerstr. 1, 81679 M\"unchen, Germany}

\author{Klaus Dolag}
\affiliation{Universit\"ats-Sternwarte, Fakult\"at f\"ur Physik, Ludwig-Maximilians-Universit\"at M\"unchen, Scheinerstr. 1, 81679 M\"unchen, Germany}
\affiliation{Max-Planck-Institut f\"ur Astrophysik, Karl-Schwarzschild-Stra{\ss}e 1, 85741 Garching, Germany}

\begin{abstract}
We train graph neural networks on halo catalogues from Gadget N-body simulations to perform field-level likelihood-free inference of cosmological parameters. The catalogues contain $\lesssim$5,000 halos with masses $\gtrsim 10^{10}~h^{-1}M_\odot$ in a periodic volume of $(25~h^{-1}{\rm Mpc})^3$; every halo in the catalogue is characterized by several properties such as position, mass, velocity, concentration, and maximum circular velocity. Our models, built to be permutationally, translationally, and rotationally invariant, do not impose a minimum scale on which to extract information and are able to infer the values of $\Omega_{\rm m}$ and $\sigma_8$ with a mean relative error of $\sim6\%$, when using positions plus velocities and positions plus masses, respectively. More importantly, we find that our models are very robust: they can infer the value of $\Omega_{\rm m}$ and $\sigma_8$ when tested using halo catalogues from thousands of N-body simulations run with five different N-body codes: Abacus, CUBEP$^3$M, Enzo, PKDGrav3, and Ramses. Surprisingly, the model trained to infer $\Omega_{\rm m}$ also works when tested on thousands of state-of-the-art CAMELS hydrodynamic simulations run with four different codes and subgrid physics implementations. Using halo properties such as concentration and maximum circular velocity allow our models to extract more information, at the expense of breaking the robustness of the models. This may happen because the different N-body codes are not converged on the relevant scales corresponding to these parameters.
\end{abstract}


\section{Introduction} 
\label{sec:intro}

%\Paco{Can you let me know the minimum halo mass in the simulations that have the same cosmology and random seed?}

To improve our knowledge on fundamental physics we need to accurately constrain the value of the parameters characterizing the composition and properties of the Universe. This can be accomplished by measuring the spatial distribution of matter and galaxies, that is sensitive to the values of the cosmological parameters. Because of this, multi-billion dollar cosmological missions such as DESI \citep{DESI_2016}, eRosita \citep{Predhel_2021}, Euclid \citep{Laureijs_2011}, PFS \citep{Takada_2014}, SKA \citep{Taylor_1999}, Roman \citep{Spergel_2015}, and Rubin \citep{LSST_2009} will be sampling the Universe at different wavelengths over gigantic cosmological volumes.

Historically, the information from cosmological observations, such as galaxy catalogues, has been compressed into a low dimensional, more manageable, data vector by using summary statistics. Theory was then used to predict the dependence of the considered summary statistics on the value of the cosmological parameters, compute the covariance matrix, and write down the likelihood of the data. Within this formalism, inference can be performed and constraints on the value of the cosmological parameters can be derived.

The above procedure is well understood and have been extensively tested. Unfortunately, it can only extract all available information if the optimal summary statistic is employed. While for Gaussian density fields this statistic is known (the power spectrum), this is not the case for non-Gaussian fields (like the spatial distribution of matter and galaxies on non-linear scales). As a result, a significant fraction of the cosmological information embedded in the data is not extracted, since the employed summary statistics are suboptimal and therefore non-lossless.

Recent advances in deep learning have enabled the possibility of using neural networks to find the optimal summary statistics for generic fields. For instance, significant work has been done on weak lensing maps from N-body simulations \citep[see e.g.][]{Schmelzle_17, Gupta_18, Ribli_19, Fluri_19, Jose_2020, Niall_2020}. Furthermore, it has been shown that neural networks can also extract information, while marginalizing over baryonic effects, on 2D maps from state-of-the-art hydrodynamic simulations \citep{Paco_2021a, Paco_2021b}. These methods not only work for 2D/3D grids but can also be applied to galaxy and halo catalogues \citep{Ntampaka_19, pablo, Makinen_2022}. 

In particular, \citet{pablo} showed that neural networks were able to infer the value of $\Omega_{\rm m}$ with a $\sim10\%$ accuracy from catalogues that contain $\sim1,000$ galaxies in a $(25~h^{-1}{\rm Mpc})^3$ volume without imposing a minimum scale. Unfortunately, their network could not be deployed with real galaxy catalogues as the model is not robust: it will only work on galaxy catalogues produced with the same simulation code that was used during training. Similar conclusions were reached by \citet{Paco_2021a} when using 2D maps from fields related to gas properties.

There could be multiple reasons behind the lack of robustness of the models: 1) galaxies/maps from  different simulations may be very distinct, 2) the network may be learning unique features associated to each subgrid model (including numerical artifacts), 3) data from the different models do not overlap due to intrinsic differences in their representations. These are complex questions that require careful attention in order to develop models that work across simulations.

In this work, we simplify the setup to quantify the impact of numerical errors intrinsic to cosmological simulations. This will allow us to quantify how much the differences in numerical approximations can propagate throughout the network and affect its robustness. For this, we will use halo catalogues from N-body simulations. We would naively expect that a network trained on halo catalogues generated from a particular N-body code would work on halo catalogues from other simulations since the different N-body codes are solving the same equations. However, different N-body codes employ different numerical schemes to solve the equations and make use of different approximations, which already lead to differences in summary statistics \citep{Heitmann_2005, Heitmann_2008, schneider, calibration_of_hmf}. Thus, it is not clear whether a field-level inference will be robust to these intrinsic differences among various codes.

We first create thousands of halo catalogues from N-body simulations run with the Gadget code \citep{Gadget} and quantify how well we can infer the value of the cosmological parameters, $\Omega_{\rm m}$ and $\sigma_{\rm 8}$, from them. Next, we test the robustness of the model on thousands of halo catalogues generated with five different N-body codes  --Abacus, CUBEP$^3$M, Enzo \citep{Enzo}, Gadget, PKDGrav \citep{PKDGrav}, and Ramses \citep{Ramses}-- and also thousands of halo catalogues from hydrodynamic simulations, run with four different codes that employ distinct subgrid physics -- Astrid \citep{Astrid1, Astrid2}, IllustrisTNG \citep{PillepichA_16a}, Magneticum \citep{Hirschmann2014}, SIMBA \citep{Dave2019_Simba}. We note that the hydrodynamic simulations not only vary the cosmology, but also the value of the astrophysical parameters that control the strength of supernova and active galactic nuclei (AGN) feedback. Finally, we also investigate the robustness of the model when several internal properties, expected to be affected by numerical errors, are used to train the model.

Our models are constructed using Graph Neural Networks (GNNs), that are trained to perform field-level likelihood-free inference. GNNs are powerful machine learning tools built to work with mathematical graphs. They are distinct in their ability to synthesize knowledge pertaining to the graph's structure and inter-node connections which allows them to leverage both global and local relations in the data. Moreover, GNNs offer many advantages such as the capability to handle irregular data structures that are built on arbitrary relations \citep{bronstein, battaglia, hamilton}, like the spatial distribution of halos. They can also be designed to exploit physical symmetries present in the data, such as the statistical isotropy and homogeneity of the Universe, without the use of data augmentation. Moreover, in contrast to convolutional neural networks, whose inputs are grid-like data that possess resolutions limited to the scales of the pixel or voxel, GNNs take in graphs which do not impose any such cut-off. This is due to the construction of the graph data in which halo clustering can be arbitrarily dense, allowing the GNNs to learn from all scales. As a result, these models can exploit the relations shared by the halos and their neighbors. In this work we follow closely the model developed by \citet{pablo}, who pioneered the usage of GNNs for extracting cosmological information from galaxy catalogues.

While GNNs have been previously used on halo catalogues from the Quijote simulations \citep{Quijote} in \citet{Makinen_2022}, there are several differences with that work. First, we concentrate on much smaller scales than \citet{Makinen_2022} ($25~h^{-1}{\rm Mpc}$ boxes versus $1000~h^{-1}{\rm Mpc}$). Second, our main goal in this work is to study robustness, and therefore we concentrate our analysis on evaluating the performance of the models on halo catalogues generated from different codes. Third, we quantify the accuracy and precision of our models, trained on halo catalogues from N-body simulations, on halo catalogues from state-of-the-art hydrodynamic simulations, enabling us to quantify the robustness of the network to hydrodynamics and astrophysical processes.

This paper is structured as follows. We first describe the data used for this project in Section \ref{sec:data}. In Section \ref{sec:methods}, we describe the architecture of our GNN models and the methods used to train and test them. In Section \ref{sec:results}, we present the results of our models. Finally, we discuss our analysis of the results in Section \ref{sec:conclusion} and summarize the main findings. 

\section{Data\label{sec:data}}

We train our models using halo catalogues from high-resolution cosmological simulations that contain the following information:

\begin{itemize}
    \item \textbf{Position}, $\vec{p}$. This is the Cartesian coordinates of the halo center in comoving-space.
    \item \textbf{Mass}, $M$. This quantity represents the total mass\footnote{In hydrodynamic simulations we include the contributions of dark matter, gas, stars, and black-holes.} of the halo.
    \item \textbf{Velocity}, $V$. This quantity represents the modulus of the 3D peculiar velocity vector.
    \item \textbf{Concentration}, $c$. The NFW halo concentration is defined as ratio between the virial radius and the scale radius of the halo: $c=R/R_s$.
    \item \textbf{Maximum circular velocity}, $V_{\rm max}$. This quantity is the maximum of the spherically-averaged rotation curve, defined as $\sqrt{GM(<R)/R}$.
\end{itemize}

The halo catalogues are generated by running \textsc{Rockstar} \citep{rockstar} on snapshots from the numerical simulations described below. Different catalogues will contain different halo properties. For instance, in some cases we may only want to use halo positions, while in others we employ positions, concentrations, and maximum circular velocities. We emphasize that each catalogue has a different number of halos. In this work we focus on halo catalogues at $z=0$.

We note that each catalogue contain less $\sim5,000$ halos and the minimum halo mass is around $10^{10}~h^{-1}M_\odot$. However, the particular number of halos and minimum mass of a given halos depends on the way the halos are chosen together with cosmology of the parent simulation.

\subsection{Simulations}

We made use of thousands of N-body and hydrodynamic simulations of volume (25 $h^{-1}$Mpc)$^3$ which have been run with 10 different codes that we briefly describe below (see Table \ref{table:sims}). All simulations follow the evolution of $256^3$ dark matter particles, plus $256^3$ gas particles in the case of the hydrodynamic simulations, from $z=127$ down to $z=0$. The initial conditions have been generated at $z=127$ using second order Lagrangian perturbation theory except for the CUBEP$^3$M simulations.

The different N-body codes follow the evolution of dark matter particles (that represent the cold dark matter plus baryonic fluid) under the effect of self-gravity using different numerical techniques and approximations. The hydrodynamic simulations solve the hydrodynamic equations using different methods but use distinct models to account for astrophysical processes such as star formation and feedback from supernova and AGN.
  
The setup is the following. We have run 6 N-body simulations that have the same initial random seed and value of the cosmological parameters ($\Omega_{\rm m}=0.3175$, $\sigma_8=0.834$) with the six different N-body codes. The rest of the N-body simulations, for each code, are organized into latin-hypercubes where the value of $\Omega_{\rm m}$, $\sigma_8$, and the initial random seed are different and vary in the range:
\begin{eqnarray}
    0.1 \leq &\Omega_{\rm m}& \leq 0.5 \ \label{eq:omegaM_range} \\
    0.6 \leq &\sigma_8& \leq 1.0~.\label{eq:sigma8_range}
\end{eqnarray}  
  
\begin{center}
\begin{table*}
\centering
	\setlength{\tabcolsep}{6pt}     
	\renewcommand{\arraystretch}{1}  
		\begin{tabular}{|c|c|c|c|c|c|}
			\hline
			Name & Type & Method & Realizations & Usage & Reference \\
			\hline
			\hline
			
			Abacus & \multirow{6}{*}{N-body}  & PP + multipole  & 51  & Testing & \citet{2021MNRAS.508..575G} \\
			
			CUBEP$^3$M &  & P$^3$M & 51 & Testing & \citet{2013MNRAS.436..540H} \\
			
			Enzo &  & AMR & 1 & Testing & \citet{Enzo}\\
			
			Gadget &  & TreePM & 1,001 & Training \& Testing & \citet{Gadget} \\
			
			PKDGrav3 &  & Tree & 1,001 & Testing & \citet{PKDGrav} \\
			
			Ramses &  & AMR & 1,001 & Testing & \citet{Ramses} \\
			
			\hline 
			Astrid & \multirow{4}{*}{Hydrodynamic} & TreePM+SPH & 1,001 & Testing & \citet{Astrid1, Astrid2}\\
			
			IllustrisTNG &   & TreePM+MMFV & 1,001 & Testing & \citet{PillepichA_16a} \\
			
			Magneticum &  & TreePM+SPH & 1 & Testing & \citet{Hirschmann2014}\\
			
			SIMBA &  & TreePM+MFM & 1,001 & Testing & \citet{Dave2019_Simba} \\
			\hline 
			
		\end{tabular}
		%}
	\caption{Characteristics of the numerical simulations used in this work. We have run 6 N-body simulations using the same initial random seed but different codes. The rest of the N-body simulations are organized into latin-hypercubes (for each code) where we vary the value of $\Omega_{\rm m}$, $\sigma_8$ and the initial random seed. Similarly, we have run 4 hydrodynamic simulations with the same cosmology and the fiducial astrophysics model using four different codes. The rest of the hydrodynamic simulations are organized into latin-hypercubes where the value of $\Omega_{\rm m}$, $\sigma_8$, four astrophysical parameters controlling supernova and AGN feedback, and the initial random seed is different.}
	\label{table:sims}
\end{table*}
\end{center}  

A similar setup is used for the hydrodynamic simulations. We have run four simulations that have the same cosmology ($\Omega_{\rm m}=0.3$, $\sigma_8=0.8$), initial random seed, and employ their fiducial subgrid physics model using four different codes. The rest of the hydrodynamic simulations, for each code, are organized into latin-hypercubes where the value of $\Omega_{\rm m}$ and $\sigma_8$ vary in the same range as the N-body simulations, but also vary four astrophysical parameters controlling the efficiency of supernova and AGN feedback. These simulations are part of the CAMELS project and we refer the reader to \citet{villaescusanavarro2020camels, CAMELS_public} for further details.
 
The simulations run with the same seed and cosmology are used to provide a qualitative measurement of the robustness of the model under a control setup, while the other simulations are used to further quantify the accuracy and precision of the models.

The six codes we use to run the N-body simulations are:
\begin{enumerate}
     \item \textbf{Abacus}. An $N$-body code which makes use of an exact near-field/far-field force decomposition. Particles pairs in the near field interact via direct pairwise evaluation, while those in the far field use a high-order multipole approximation. The resulting force errors are tiny, with median fractional error $\mathcal{O}(10^{-5})$. Leap frog integration is performed with an global time step whose size is determined at each time step. The time step parameter, \texttt{TimeStepAccel}, was chosen to be $0.15$ instead of the usual $0.25$ for increased accuracy. The resulting time step is very small compared to dynamical times outside of cluster cores, as the simulations take about 2,000 steps to $z=0$.  Spline softening was used, with a softening length fixed in proper coordinates to a Plummer-equivalent length of $\ell/40$, capped at $0.3\ell$ in comoving coordinates. The code is described in described in \citet{2021MNRAS.508..575G}; the mathematical method in \citet{2009PhDT.......175M}; and the softening scheme is validated in \citet{2021MNRAS.504.3550G}. We have run 51 simulations with Abacus: 1 simulation with a shared cosmology and initial random seed among codes and 50 simulations in a latin-hypercube with varying values of $\Omega_{\rm m}$ and $\sigma_8$.
    
    \item \textbf{CUBEP$^3$M} A particle-particle particle-mesh (P$^3$M) code described in \citet{2013MNRAS.436..540H} where the gravitational force over large distances is obtained using a two-level particle mesh calculation, while subgrid resolution is obtained via direct particle-particle calculations.  We used the high accuracy parameters for the force calculation tested in \citet{PhysRevD.100.083528}, but furthermore found that the code ran faster and yielded better results when using a higher ratio of grid cells to particles (64:1 instead of 8:1) at fixed softening length ($4.9$ kpc/h). We have run 51 simulations with CUBEP$^3$M: 1 simulation with shared cosmology and initial random seed among codes and 50 simulations in a latin-hypercube. For the simulation sharing the cosmology and initial random seed we used the exact same initial particles as in the other codes, whereas the CUBEP$^3$M initial conditions, generated using the Zeldovich approximation, was used for the 50 simulations in the latin-hypercube.
    
    \item \textbf{Enzo}. An Adaptative Mesh Refinement (AMR) code described in \citet{Enzo} that uses a fast Fourier technique \citep{Hockney_1998} to solve Poisson's equation on the root grid and a multigrid technique on the subgrids. The dark matter particles are evolved using a kick-drift algorithm that provides second order accuracy. We refine a cell when the dark matter density in it reaches above $3\times 2^{3l}$ times the background density on the root grid where $l$ is the refinement level. In our simulation we include 7 refinement levels. We also set the CourantSafetyNumber and ParticleCourantSafetyNumber parameters to 0.15 and 0.125 respectively to ensure smaller time steps and that the particle displacements are smaller relative to the size of the most refined cell, so as to have particle evolution accurate and converged. We only have one simulation for this code, whose cosmology and initial random seed is shared among codes.
    
    \item \textbf{Gadget}. A TreePM code described in \citet{Gadget}. We have used Gadget-III for these simulations, and these simulations are also part of the CAMELS project. The PM grid contains $512^3$ voxels, and the FFT are computed using double precision. We set the value of the relevant parameters to ErrTolIntAccuracy=0.025, MaxSizeTimestep=0.005, ErrTolTheta=0.5, ErrTolForceAcc=0.005, TreeDomainUpdateFrequency=0.01. The value of the softening length is set to 1/40 of the mean inter-particle distance. We have 1,001 of these simulations: 1 simulation with shared cosmology and initial random seed among codes and 1,000 simulations that have different values of $\Omega_{\rm m}$, $\sigma_8$, and initial random seed. We use the halo catalogues from these simulations to train the models.
    
    \item \textbf{PKDGrav3}. Described in \citet{PKDGrav}, it computes the forces between particles using a highly performing and memory efficient version of the Fast Multipole Method (FMM, \citealt{Greengard1987}), with typical run-times that scale linearly with the number of particles.  The FMM algorithm is implemented using a binary tree structure, which reduces the number of terms of the multipole expansion needed to evaluate the force on the particles. PKDGRAV3 run-time can be further improved by using GPU accelerated nodes. GPU acceleration is used to evaluate particles-particles, particles-cells interactions and periodic boundary conditions (implemented using the Ewald summation method), whereas standard CPUs are used to build and walk the trees. We have run 1,001 N-body simulations with PKDGrav3: 1 simulation with shared cosmology and initial random seed among codes and 1,000 simulations with different values of $\Omega_{\rm m}$, $\sigma_8$ and initial random seed that are organized in a latin-hypercube. 
    
    \item \textbf{Ramses}. An N-body code based on the Adaptive Particle Mesh technique described in \citet{Ramses}. The Adaptive Mesh Refinement (AMR) framework is based on a graded octree for which cells are individually refined when the mass exceeds 8 times the particle mass. Mass deposition is performed using the Cloud-In-Cell scheme. Poisson's equation is solved level by level using Dirichlet boundary conditions from the coarser level and a Multigrid relaxation solver. Time integration is performed using the Verlet algorithm (aka adaptive leap frog). The minimum level of refinement was set to $\ell_{\rm min}=8$, corresponding to 256$^3$ base grid cells (one per particle on average). The maximum level of refinement was found to be $\ell_{\rm max}=15$, corresponding to a minimum cell size of roughly 1~h$^{-1}$kpc. We have run 1,001 simulations with Ramses: 1 simulation with shared cosmology and initial random seed among codes, and 1,000 simulations with different values of $\Omega_{\rm m}$, $\sigma_8$, and initial random seed that are organized in a latin-hypercube.
\end{enumerate}

The hydrodynamic simulations have been run with the codes MP-Gadget, Arepo, OpenGadget, and Gizmo. Each of codes employs a different subgrid model. In these simulations, that are part of the CAMELS project \citep{villaescusanavarro2020camels}, we vary the value of $\Omega_{\rm m}$, $\sigma_8$, the initial random seed, and four astrophysical parameters that control the efficiency of supernova and AGN feedback. Instead of referring to these simulations by the name of the code used to run them, we will call them by name of the flagship simulations associated to them and their subgrid model, i.e. ASTRID, IllustrisTNG, Magneticum, and SIMBA, respectively. We now briefly describe the simulations from the different codes:

\begin{enumerate}
 \setcounter{enumi}{6}
    \item \textbf{ASTRID}. These simulations have been run with MP-Gadget simulation code, a massively scalable version of the code P-Gadget3 \citep{Gadget}, to solve the gravity (with TreePM), hydrodynamics (with pressure-entropy formulation of SPH), and astrophysical processes with a series of subgrid models as employed in the ASTRID simulation \citep{Astrid1, Astrid2}. We made use of 1,001 simulations. One simulation was run with shared cosmology and initial random seed among hydrodynamic codes (and a fiducial value of the astrophysical parameters). The other 1,000 simulations have different values of $\Omega_{\rm m}$, $\sigma_8$, four astrophysical parameters and the initial random seed, that are organized in a latin-hypercube.
    \item \textbf{IllustrisTNG}. These simulations have been run with the AREPO code \cite{Arepo, Arepo_public}, making use of a TreePM plus moving-mesh finite volume (MMFV) method, and employing the same subgrid model as the IllustrisTNG simulation \citep{WeinbergerR_16a,PillepichA_16a}. We made use of 1,001 simulations. One simulation was run with shared cosmology and initial random seed among hydrodynamic codes (and a fiducial value of the astrophysical parameters). The other 1,000 simulations have different values of $\Omega_{\rm m}$, $\sigma_8$, four astrophysical parameters and the initial random seed, that are organized in a latin-hypercube.
    \item \textbf{Magneticum}. We have one simulation in this category, that has been run with the code OpenGadget3, which is an advanced version of the code P-Gadget3, the developers version of P-Gadget2 \citep{Gadget}. The code adopts a modern state of the art SPH-scheme following the implementation of \citet{Beck2016} with a Wendland C4 kernel that adopts 200 neighbours as well as an physical treatment for thermal conduction following \citep[][]{Dolag2004, Jubelgas2004}. Furthermore, we adopted the models for star formation and feedback used in the \textsc{magneticum} simulations as outlined by \citet{Hirschmann2014}. The simulation has the same value of $\Omega_{\rm m}$, $\sigma_8$, and the initial random seed than their Astrid, IllustrisTNG, and SIMBA counterpart. The value of the astrophysical parameters is set to its fiducial value.

    \item \textbf{SIMBA}. These simulations have been run with the GIZMO code \citep{Gizmo} with a TreePM plus mesh-free finite mass method (MFM), employing the same subgrid model as the SIMBA simulation \citep{Dave2019_Simba}.  We made use of 1,001 simulations. One simulation was run with shared cosmology and initial random seed among hydrodynamic codes (and a fiducial value of the astrophysical parameters). The other 1,000 simulations have different values of $\Omega_{\rm m}$, $\sigma_8$, four astrophysical parameters and the initial random seed, that are organized in a latin-hypercube.
\end{enumerate}

\begin{comment}
\begin{table}
	\begin{center}
	%\resizebox{0.5\textwidth}{!}{
	\setlength{\tabcolsep}{10pt}    % row separation
	\renewcommand{\arraystretch}{1.3}   % column separation
		\begin{tabular}{|c|c|c|}
			\hline
			Name & Type & Realizations \\
			\hline
			\hline
			Abacus & N-body & 50 \\
			\hline
			CUBEP$^3$M & N-body & 50 \\
			\hline 
			Enzo & N-body & 1 \\
			\hline 
			Gadget & N-body & 1,000 \\
			\hline 
			PKDGrav & N-body & 1,000 \\
			\hline 
			Ramses & N-body & 1,000 \\
			\hline 
			IllustrisTNG & hydrodynamic & 1,000 \\
			\hline 
			SIMBA & hydrodynamic & 1,000 \\
			\hline 
			Astrid & hydrodynamic & 1,000 \\
			\hline 
			Magneticum & hydrodynamic & 1 \\
			\hline 
			
		\end{tabular}
		%}
	\end{center}
	\caption{Characteristics of the numerical simulations used in this work.}
	\label{table:sims}
\end{table}

\end{comment}

%%%%%%%%%%%%%%%%%%%%%%%%%%%%%%%%%%%%%%%%%%%%%%%
%%%%%%%%%%%%%%%%%%%%%%%%%%%%%%%%%%%%%%%%%%%%%%%
\section{Methods} \label{sec:methods}

In this section we describe 1) how we create the input to the mode (cosmic graphs), 2) the architecture of the GNN, 3) the training procedure, and 4) the metrics we use to evaluate the accuracy and precision of the model.

\subsection{Model input: Halo Graphs\label{subsubsec:graphs}}

The input of our models is a mathematical graph, defined by the tuple $\mathcal{G} = (\mathcal{V}, \mathcal{E})$, where $\mathcal{V}$ is the set of nodes and $\mathcal{E}$ is the set of edges that connect the nodes. We refer the reader to \cite{battaglia} for more details on these mathematical definitions in the context of deep learning.

Following \cite{pablo}, we construct graphs from the halo catalogues as follows. The nodes  represent the halos and an edge is established between two nodes if their distance is smaller than the \textit{linking radius}, $r$, a hyperparameter of our model that is optimized during training (as explained later). Thus, two nodes $i,j \in \mathcal{V}$ are referred to as neighbors if they are connected via an edge, $(i,j) \in  \mathcal{E}$. Note that we do not consider self-loops, i.e. a node is not connected to itself. Moreover, we account for periodic boundary conditions when computing distances between nodes.

The nodes and the edges can have different properties associated to them, that we denote as  $\textbf{h}_i^{(n)}$ and $\textbf{e}_{ij}^{(n)}$, respectively. As explained in the later sections, the architecture of the GNNs consists of multiple layers that take a graph as the input and outputs an updated graph. For this reason, we denote the node and edge features at the $n^{th}$ layer with the superscript $n$.

We train different models that make use of graphs with different node properties to infer $\Omega_{\rm m}$, $\sigma_8$ or both parameters. The initial node features, represented by $\textbf{h}_i^{(0)}$, that we use are:
\begin{itemize}
\setlength\itemsep{-0.5em}
    \item the halo velocity modulus, $V$, when inferring $\Omega_{\rm m}$
    \item the halo mass, $M$, when inferring $\sigma_8$
    \item the halo concentration, $c$, and maximum circular velocity, $V_{\rm max}$, when inferring both $\Omega_{\rm m}$ and $\sigma_8$.
\end{itemize}
We explain the reasoning for the choice of input node features in Section \ref{sec:results}.

The edge features between nodes $i$ and $j$ at the $n^{th}$ layer are represented by $\textbf{e}_{ij}^{(n)}$ and they contain information about the spatial distribution of halos. We build our models to be translational and rotational invariant. These symmetries can be imposed by requiring that the output graph is invariant under this rigid transformation of $\vec{p}_i$: $\vec{p}_i \rightarrow \textbf{R}\vec{p}_i+\textbf{T}$. Here, $\vec{p}_i$ is the position of the halo at node $i$, and $\textbf{R}$ and $\textbf{T}$ are rotation and translation matrices, respectively. We follow \citet{pablo} and impose rotational and translation symmetry by using as initial edge features the vector $\textbf{e}_{ij}^{(0)}=[\alpha_{ij}, \beta_{ij}, \gamma_{ij}]$, where $[...]$ denotes a concatenation along the features axis and the three edge features are defined as:

\begin{eqnarray}    
\alpha_{ij} &&= \frac{\vec{p}_i-\bar{p}}{|\vec{p}_i-\bar{p}|} \cdot \frac{\vec{p}_j-\bar{p}}{|\vec{p}_j-\bar{p}|}\label{Eq:edge_features1}\\
\beta_{ij} &&= \frac{\vec{p}_i-\bar{p}}{|\vec{p}_i-\bar{p}|} \cdot \frac{d_{ij}}{|d_{ij}|}\label{Eq:edge_features2}\\
\gamma_{ij}&&=\frac{|d_{ij}|}{r}\label{Eq:edge_features3}, 
\end{eqnarray}

with $d_{ij}=\vec{p}_i-\vec{p}_j$ being the relative distance between the nodes $i$ and $j$ and $\bar{p}$ is the centroid\footnote{Note that other choices of reference position vector would also work, like the center of mass, as long as translational invariance is preserved.} of the halo distribution. These edge features arise from taking scalar products between the different vectors involved $\vec{p}_i-\bar{p}$, $\vec{p}_j-\bar{p}$, and $\vec{p}_i-\vec{p}_j$. $\alpha_{ij}$ describes the angle between the vectors defining the position of node $i$ and its neighbor node $j$, while $\beta_{ij}$ describes the angle between the vectors defining the position of node $i$ and the separation between nodes $i$ and $j$. Note that we have normalized the distance, $d_{ij}$, by dividing it with the linking radius, $r$, to have dimensionless edge features. These three scalars contain the same information as the vector with the relative position of two nodes, that could be used as edge features. Since translations and rotations will not affect the values of these edge features, our model will automatically satisfy these symmetries. We refer the reader to \cite{pablo} for more details on this construction.

\subsection{Architecture\label{subsubsec:architetcure}}

For the construction of our GNNs, we follow closely the \textsc{CosmoGraphNet}\footnote{\url{https://github.com/PabloVD/CosmoGraphNet}} model architecture \citep{pablo_villanueva_domingo_2022_6485804}, presented in \cite{pablo}. The predominant structure underlying the GNN is a composition of $N$ message passing layers with a final aggregation layer. The messages are constructed by encoding the input node and edge features with multilayer perceptrons (MLP). The message passing scheme is a recursive process of exchanging and aggregating messages between each node's neighbors and edges, and iteratively updating them. This creates hidden feature vectors that are ultimately used to make the prediction of the graph's global property, which is the desired target parameter. 

Each message passing layer takes a graph as an input and outputs another graph with updated node and edge features. To achieve this, each layer consists of both edge and node models that update the respective features of the input graph. At the $n^{th}$ layer, the input to the edge model are the features of the node $i$, the neighboring node $j$, and their shared edge. This information is passed through a MLP, denoted by $\phi ^e$ and the output are the updated edge features:

\begin{equation}
    \textbf{e}_{ij}^{(n+1)}=\phi^e\left(\left[\textbf{h}_i^{(n)},\textbf{h}_j^{(n)},\textbf{e}_{ij}^{(n)}\right]\right). 
    \label{eq:edgelayer}
\end{equation}

Meanwhile, the input to the node model are the feature node $i$ and the updated edge feature, which are propogated through another MLP, denoted by $\phi ^h$, to output the updated node features

\begin{equation}
    \textbf{h}_i^{(n+1)} = \phi^h \left(\left[\textbf{h}_i^{(n)}, \bigoplus_{j \in \mathcal{N}_i} \textbf{e}_{ij}^{(n+1)}\right]\right)~.
    \label{eq:nodelayer}
\end{equation}

Here, we use a permutationally invariant aggregation function denoted by $\bigoplus$ to aggregate the node features of the neighbor nodes $j \in \mathcal{N}_i$ which are connected via an edge to node $i$. As in \citet{pablo}, this function is a concatenation of the maximum, sum, and mean operators:

\begin{equation}
    \bigoplus_{j \in \mathcal{N}_i} \textbf{e}_{ij}^{(n+1)}=\left[\max_{j \in \mathcal{N}_i} \textbf{e}_{ij}^{(n+1)}, \sum_{j \in \mathcal{N}_i} \textbf{e}_{ij}^{(n+1)}, \frac{ \sum_{j \in \mathcal{N}_j} \textbf{e}_{ij}^{(n+1)}} {\sum_{j \in \mathcal{N}_j}} \right]
    \label{eq:pooling}
\end{equation}

In cases where we exclude the internal properties of halos in the construction of our graphs and only work with their relative positions, we do not input initial node features. As a result, we include an initial layer of edge and node models to update the respective features:
\begin{eqnarray}
    \textbf{e}_{ij}^{(1)}&=&\phi^e \left(\textbf{e}_{ij}^{(0)}\right) \\
    \textbf{h}_i^{(1)} &=& \phi^h \left(\bigoplus_{j \in \mathcal{N}_i} \textbf{e}_{ij}^{(1)}\right)~,
\end{eqnarray}
and the subsequent layers follow Equations \ref{eq:edgelayer} and \ref{eq:nodelayer}.

For all models, the final layer in the architecture aggregates the node features output by the $N^{th}$ message passing layer to construct the prediction $\textbf{y}$,

\begin{equation}
    \textbf{y} = \phi ^u \left( \left[\bigoplus_{i \in \mathcal{G}} \textbf{h}_i^{(N)}\right] \right),
    \label{Eq:GNN_output}
\end{equation}

where $\bigoplus_{i \in \mathcal{G}}$ operates over all nodes in the graph and $\phi ^u$ is another MLP that extracts the target information. To emphasize, the model architecture imposes permutation invariance in the use of the aggregation function $\bigoplus$ so that the ordering of the nodes do not affect the output of the model \citep{bronstein}. This is in addition to the translational and rotational symmetries imposed by the definition of the edge features. 

\subsubsection{Loss Function \label{subsubsec:loss}}
To perform likelihood-free inference we train the GNNs to predict the posterior mean $\mu_i$ and standard deviation $\sigma_i$ for the considered cosmological parameter $\theta_i$ (note that this can also represent several parameters, not just one). Thus, the output of the model is $\mathbf{y} = [\mu_i, \sigma_i]$, where
\begin{eqnarray}
    \mu_i(\mathcal{G}) &=& \int_{\theta_i} p(\theta_i|\mathcal{G}) \theta_i d\theta_i~,\\
    \sigma_i^2(\mathcal{G}) &=& \int_{\theta_i} p(\theta_i|\mathcal{G}) (\theta_i - \mu_i)^2 d\theta_i~.
\end{eqnarray}

Here, $\mathcal{G}$ represents the graph and $p(\theta_i|\mathcal{G})$ is the marginal posterior over the parameter $\theta_i$

\begin{equation}
    p(\theta_i|\mathcal{G}) = \int_{\theta_i} p(\theta_1, \theta_2, ... \theta_n|\mathcal{G})d\theta_1d\theta_2...d\theta_n
\end{equation}

These predictions can be achieved by optimizing the loss function of the form discussed in \citet{jeffrey_wandelt} 
\begin{equation}
\begin{split}
    \mathcal{L} & = \log{\bigg(\sum_{j\in batch}(\theta_{i,j} - \mu_{i,j})\bigg)^2} + \\
    & \log{\bigg(\sum_{j\in batch}\big((\theta_{i,j} - \mu_{i,j})^2 - \sigma_{i,j}^2\big)\bigg)^2}
\end{split}
\end{equation}

where the sums are performed over the halo catalogues in the batch. Note that we take the log of the sums over the batches to ensure that both terms in the loss function have the same order of magnitude. Further details on this can be found in \cite{CMD}.

\subsection{Training procedure}\label{subsec:training_procedure}

We train and test the models using graphs constructed from halo catalogues of the different simulations. The networks are trained using halo catalogues from Gadget simulations and tested using catalogues from all simulations. For Gadget, we split the simulations into training (80\%), validation (10\%), and testing (10\%) data sets before creating halo catalogues for each simulation. For the other codes we use the entirety of the dataset for testing. Note that the simulations that share the same initial random seed and cosmology are always used for testing only.

For each simulation we generate 20 catalogues by taking all halos with masses larger than $M_{\rm X}$, where $M_{\rm X}$ is a randomly chosen number between between $100\,m_{\rm p}$ and $700\,m_{\rm p}$. Here, $m_p$ is the mass of a single dark matter particle\footnote{Note that while $m_p$ is well-defined in the case of N-body simulations, for hydrodynamic simulations we consider it to be the effective particle mass, defined as $m_p = \frac{1}{N_c}\Omega_{\rm m}V\rho_c$, where $V$ is the volume of the simulation, $\rho_c$ is the Universe's critical density today, and $N_c=256^3$ is the effective number of particles.}. We note that the resulting catalogues contain less than $\sim 5,000$ halos. 

We follow this procedure so that the network learns to make predictions independently of the number density of halos a particular catalogue contains. We find this procedure to be crucial to make our models robust\footnote{We initially trained the networks using a single catalogue per simulation chosen with a given criterion and find that our model was not robust.}. 

We standardize the values of input node features as
\begin{equation}
    \tilde{x}=\frac{x - {\mu}}{{\delta}},
\end{equation}
where $\mu$ and $\delta$ denote the mean and standard deviation of the feature $x$. In catalogues containing halo mass, we use $M\longrightarrow\log_{10}(M)$ to accommodate the large dynamical range involved. We also normalize the values of the target cosmological parameters using:

\begin{equation}
    \theta_i=\frac{\theta_i - \theta_{min}}{\theta_{max} - \theta_{min}},
\end{equation}

where $\theta_{min}$ and $\theta_{max}$ are the minimum and maximum values of the ranges of the cosmological parameter, $\theta_i$, as listed in Eq.~\ref{eq:omegaM_range}, \ref{eq:sigma8_range}.

Our model is implemented in PyTorch \citep{Pytorch} and PyTorch Geometric \citep{Fey_Fast_Graph_Representation_2019}. We use the AdamW optimizer \citep{AdamW} with beta values equal to 0.9 and 0.999. We train the network using a batch size of 8 for 500 epochs. The hyperparameters for our model are 1) the learning rate, 2) the weight decay, 3) the linking radius, 4) the number of message passing layers, and 5) the number of hidden features per layer. We use the \textsc{optuna} code \citep{Optuna} to perform Bayesian optimization and find the best-value of these hyper-parameters for each model. For each model we run 100 trials, where each trial consists of training the model using selected values of the hyper-parameters. We perform the optimization of the hyper-parameters requiring to achieve the lowest validation loss possible.

Once trained, we test the models on halo catalogues from all N-body and hydrodynamic, simulations. 

\subsection{Accuracy Metrics\label{acc_metrics}}

For the graph $i$, with true value of the considered parameter $y_{\rm truth,i}$, our models output the posterior mean, $y_{\rm infer,i}$, and standard deviation $\sigma_i$. To evaluate the performance of our models, we follow \cite{pablo} and adopt four different metrics:

\begin{enumerate}
    \item \textbf{Mean relative error},  $\epsilon$, defined as
    \begin{equation}
    \epsilon = \frac{1}{N} \sum_i^N \frac{|y_{{\rm truth}, i} - y_{{\rm infer}, i}|}{y_{{\rm truth}, i}},
\end{equation}
where $N$ is the number of halo catalogues in the test set.

    \item \textbf{Coefficient of determination}, $R^2$, defined as
    \begin{equation}
    R^2 = 1 - \frac{\sum_i^N (y_{{\rm truth}, i} - y_{{\rm infer}, i})^2}{\sum_i^N (y_{{\rm truth}, i} - \overline{y}_{{\rm truth}})^2},
    \end{equation}
    
    \item \textbf{Root mean squared error}, RMSE, defined as:
    \begin{equation}
    {\rm RMSE} = \sqrt{\frac{1}{N}\sum_{i=1}^N \left(y_{{\rm truth}, i} - y_{\rm infer}\right)^2}
    \end{equation}

    \item \textbf{Chi squared}, $\chi^2$, defined as:
    \begin{equation}
\chi^2=\frac{1}{N}\sum_{i=1}^N \frac{(y_{\rm truth,i} - y_{\rm infer,i})^2}{\sigma_{i}^2}~.
\label{Eq:chi2}
\end{equation} 

Note that a value of $\chi^2$ is close to 1 suggests that the standard deviations are accurately predicted. On the other hand, a larger or lower value indicates that the uncertainties are under- or overestimated, respectively. 
\end{enumerate}

Note that the sums in all expressions above run over the graphs in the test set. When doing multiparameter inference (i.e. when the model predicts both $\Omega_{\rm m}$ and $\sigma_8$), the above expressions are used for each parameter.

%%%%%%%%%%%%%%%%%%%%%%%%%%%%%%%%%%%%%%%%%%%%%%%%%
%%%%%%%%%%%%%%%%%%%%%%%%%%%%%%%%%%%%%%%%%%%%%%%%%
\section{Results} 
\label{sec:results}

In this section we show the results we obtain from training our networks. We have trained different models that use catalogues containing different halo properties to infer either $\Omega_{\rm m}$, $\sigma_8$, or both. Table \ref{tab:models} summarizes the results of the different networks trained.

We note that the performance metrics that we report in this section measure the aleatoric errors of the models which quantify the uncertainty of the predictions due to the variability in the data (i.e. due to cosmic variance). This is because we find the epistemic errors (the ones quantifying the errors associated to the network itself, not the data) to be much smaller than the aleatoric ones. We provide further details in the Appendix \ref{sec:epistemic}.

\subsection{Inferring \texorpdfstring{$\Omega_{\rm m}$}{Om} \label{subsec:inferring_omegaM}}

We start by training and testing a GNN to infer the parameter $\Omega_{\rm m}$ using catalogues that only contain the positions of the halos. In this case, the nodes do not contain any information and only the edges do. When we evaluate this model on the test set of the Gadget simulations, we find that the GNN is able to infer the value of $\Omega_{\rm m}$ with a mean relative error of $\epsilon = 10.1$ \% and a coefficient of determination of $R^2 = 0.91$. Additionally, the Chi-squared value is $\chi ^2 = 1.14$ which indicates that the posterior standard deviations are predicted accurately. To investigate whether our field-level inference is extracting more information than traditional summary statistics we train a MLP to perform likelihood-free inference for $\Omega_{\rm m}$ from the halo power spectrum for $k<30~h{\rm Mpc}^{-1}$, computed from catalogues of the Gadget simulations. We find that the model obtains a mean relative error of $\epsilon = 18.3$ \%, with $\chi ^2 = 2.30$, indicating that the GNN is able to extract much more information at the field level.
  
\begin{figure*}
    \centering
    \includegraphics[width=0.99\textwidth]{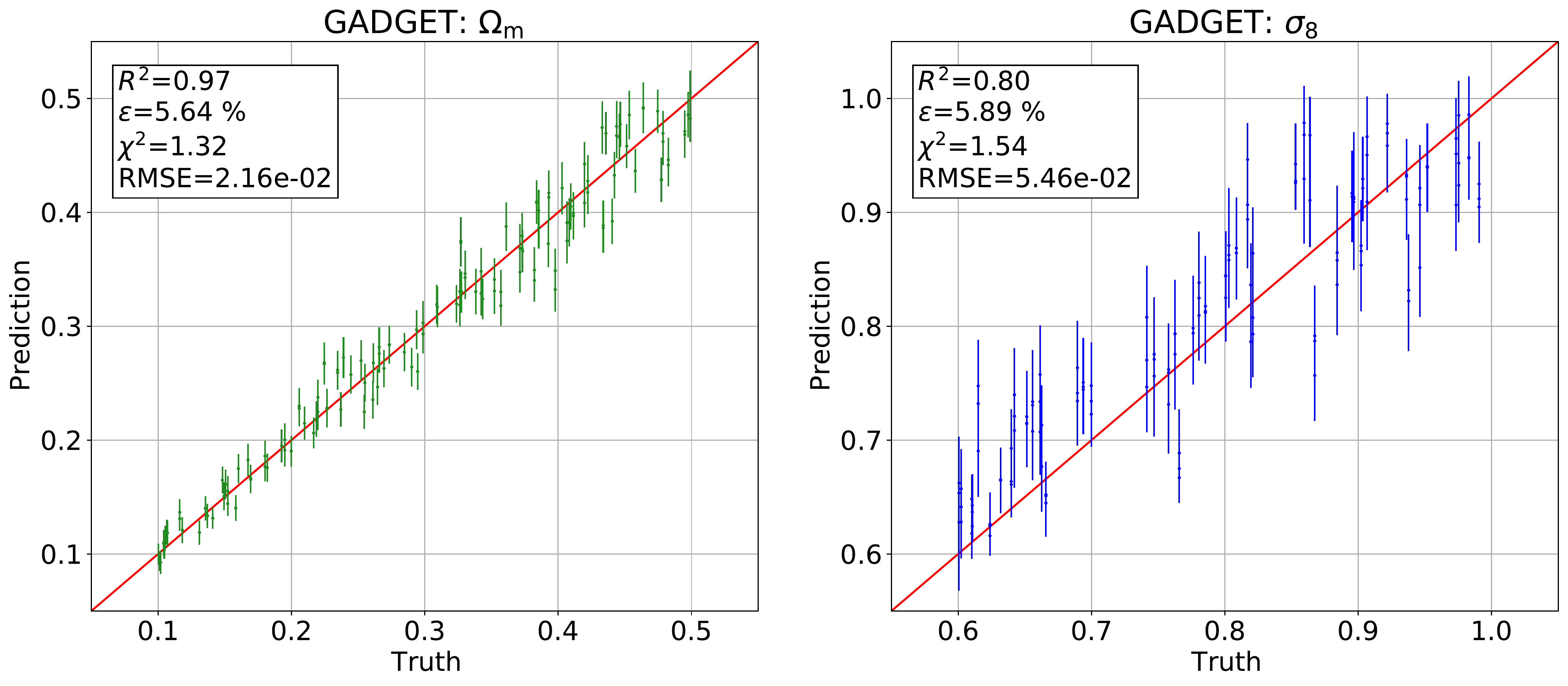}
    \caption{We train two GNNs to perform likelihood-free inference for the cosmological parameters $\Omega_{\rm m}$ (left) and $\sigma_{\rm 8}$ (right), respectively. To infer $\Omega_{\rm m}$ ($\sigma_8$), the halo catalogues contain the velocity modulus, $V$ (halo mass, $M$), as node feature. For both models, we use the relative halo positions as the initial edge features, defined in Equations \ref{Eq:edge_features1}, \ref{Eq:edge_features2}, \ref{Eq:edge_features3}. The models are trained using halo catalogues from the Gadget simulations. For each simulation, we generate 20 graphs by taken all halos with masses above a random threshold (see text for details). Once the networks are trained we evaluate them on halo catalogues from the Gadget test set. As can be seen, the models are able to infer the correct value of the cosmological parameters with a $\sim 6\%$ accuracy.}
    \label{fig:gadget_individual}
\end{figure*}  
  
Next, we train models using catalogues that not only contain halo positions but also the modulus of their peculiar velocities, $V$. We do this because we expect that including other halo properties will increase the amount of information available. For simplicity and to preserve the symmetries that we impose in the graphs, we take the modulus of the 3D velocity vector. Note that we do not include the halo mass, $M$, as an input feature because there is an underlying linear relationship between $\Omega_{\rm m}$ and the minimum $M$ of each catalogue, via $M_{\rm min}=N_pm_p$, where $N_p$ is the number of dark matter particles. Should we have trained the model using a single threshold, we would have gotten artificially tight constraints on $\Omega_{\rm m}$. To avoid the existence of any leakage of information from $M_X$ to $\Omega_{\rm m}$ we simply not include halo mass as a node feature.

\begin{figure*}
    \centering
    \includegraphics[width=1\textwidth]{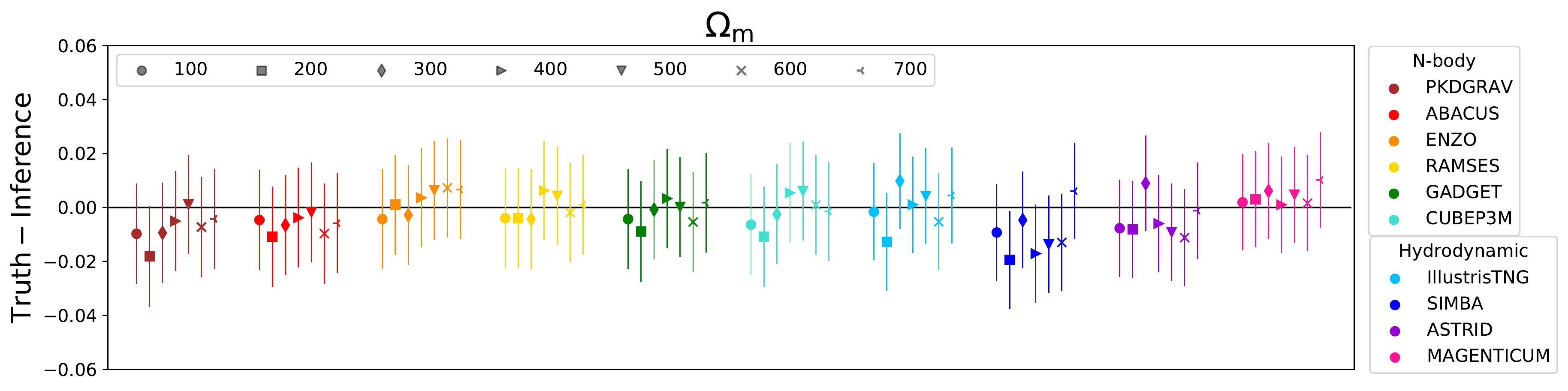}
    \includegraphics[width=1\textwidth]{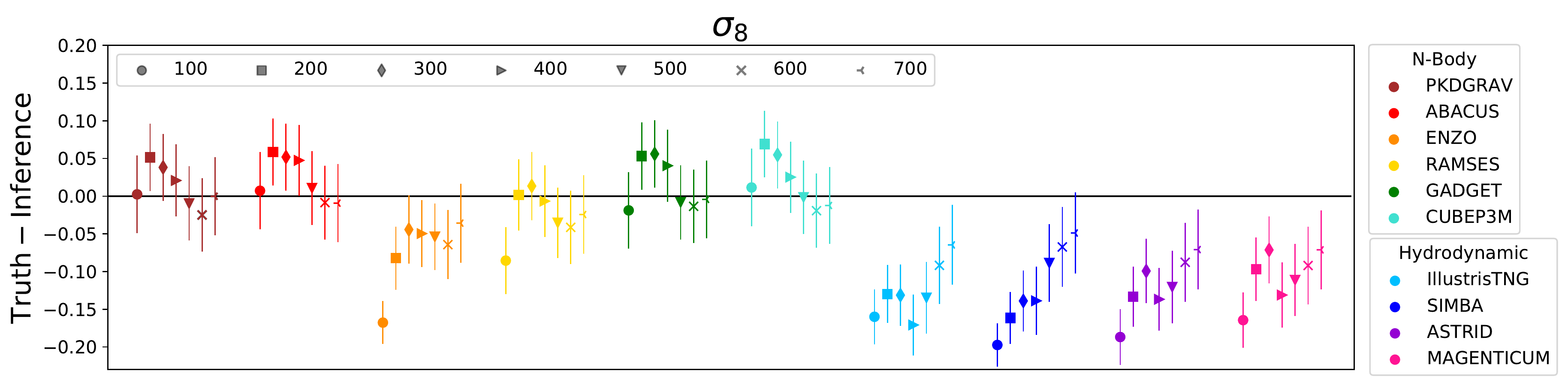}
    \caption{\textbf{Top:} We train a rotationally, translationally, and permutationally invariant GNN to perform likelihood-free inference for the cosmological parameter $\Omega_{\rm m}$. The input to the model are halo catalogues from Gadget that only carry information about halo positions and peculiar velocity moduli. Once trained, we test the model on halo catalogues from different N-body and hydrodynamic simulations as indicated in the legend.  We note that simulations of the same type, either N-body or hydrodynamic, are run with the same initial conditions, cosmology (and fiducial astrophysics for the hydrodynamic simulations). For each simulation we generate 7 catalogues. Each halo catalogue contains all halos with masses above $Nm_p$, where $m_p$ is the particle mass and $N$ can be 100, 200, 300, 400, 500, 600, or 700 (see legend). The y-axis represents the difference between the truth and the inference. As it can be seen, this model exhibits surprising extrapolation properties and is robust to all simulation codes. \textbf{Bottom:} Same as above but for a network trained to infer $\sigma_{\rm 8}$ from catalogues that contain halo positions and masses. As it can be seen, this model is robust only for N-body halos that contain at least 300 particles. Moreover, the predictions exhibit poor accuracy for all hydrodynamic simulations, suggesting that this model is more sensitive to the additional astrophysical and baryonic effects than the one for $\Omega_{\rm m}$.}
    \label{fig:same_seed}
\end{figure*}

We show the results of testing the model on Gadget catalogues in Fig. \ref{fig:gadget_individual}. As can be seen, the constraints are significantly better: the mean relative error decreases to 5.6 \% while the coefficient of determination increased to $R^2 = 0.97$. The Chi squared value is $\chi ^2 = 1.32$.

\subsubsection{Clustering}

We ask ourselves whether the information the network is extracting comes from the clustering of halos or perhaps just from the velocities themselves. To address this, we train a DeepSet model \citep{deepset} to perform likelihood-free inference on the value of $\Omega_{\rm m}$. In contrast to GNNs, DeepSets take as the input a set of nodes that are not connected via edges and do not employ the message passing scheme to leverage the relations between elements of the set. Moreover, this model does not receive any information about the halo positions and thus cannot extract any information from the clustering of the halos. However, in order to make the prediction, the DeepSet model still follows the permutationally invariant operation seen in Eq.~\ref{Eq:GNN_output} for aggregating the node features, which are the halo velocity moduli.

When evaluated on the test set, this model performs with a relative error of $\epsilon = 17.4$ \%, which is a noticeable increase from the models trained using catalogues that include positions and positions plus velocities. This implies that the halo velocity modulus alone does not provide sufficient information and most of the information used by the GNNs to predict $\Omega_{\rm m}$ comes from the clustering of the halos. Hence, going forward, we will only report the results of the GNNs trained using both the halo positions and velocity moduli.

\subsubsection{Robustness}

Next, we study the robustness of our model by testing it on halo catalogues produced by different N-body codes. We emphasize the importance of this test, as field-level inference is expected to work under this control setup where different codes are solving the same equations using just different numerical techniques. If this is not the case, that would mean that numerical errors from the simulations are propagating into the network and perhaps the network is learning unique features associated to each simulation.

To check this, we test the models trained on halo catalogues from Gadget on halo catalogues from Abacus, Ramses, PKDGrav3, Enzo, and CUBEP$^3$M. We find that the model is able to accurately infer the value of $\Omega_{\rm m}$ on all halo catalogues from the different simulations. In Fig. \ref{fig:same_seed} we show the constraints derived by the model from halo catalogues obtained from the simulations that share the same cosmology and initial conditions, but different N-body codes. As it can be seen, the model performs with similar accuracy for all simulations, demonstrating its robustness. We present more detailed results of this test in Appendix \ref{sec:additional plots}, where the figures depict the accuracies of the model when tested on 50 catalogues of different cosmologies from each simulation code. 

We emphasize that this is not a trivial test. The catalogues from the different simulations, even if having the same initial conditions, will have different number of halos and the halos in common will have slightly different positions, and other properties, due to the numerical approximations of the different codes. It is important to remark that a key ingredient to achieve this robustness was to generate many catalogues with different number densities from the same simulation. In that case, the network was trained to make predictions independently of the number density\footnote{This is particularly important for the Enzo and Ramses simulations, where the abundance of low mass halos is significantly lower than that from the other codes.}.

Next, we test the model on halo catalogues from different hydrodynamic simulations: Astrid, IllustrisTNG, Magneticum, and SIMBA. Surprisingly, we find that the model works also when tested on these halo catalogues. In Fig. \ref{fig:same_seed} we show the results when the model is tested on the simulations that share the same cosmology and initial random seed (these simulations have a fiducial value of the astrophysical parameters). As can be seen, the precision and accuracy of the model remains comparable to the one obtained from the catalogues from the N-body codes, indicating that the model is robust even to hydrodynamics, changes in astrophysical parameters, and subgrid physics models. We emphasize that this is a not trivial result at all, since the positions and properties of halos are expected to be affected by hydrodynamics and astrophysical processes. This result may suggest that the model is exploiting a fundamental relation between the clustering of the halos, the halo velocities, and $\Omega_{\rm m}$ that is robust to cosmology and astrophysics.

\begin{table*}[t]
\begin{center}
\begin{tabular}{|c|c|c|c|c|}
\hline
\textbf{Parameter(s) Inferred} & \textbf{Accuracy} ($\epsilon$) & \textbf{Node Features} &\textbf{Edge Features} & \textbf{Robust} \\
\hline \hline 
$\Omega_{\rm m}$ & 10.1 \% & - & $\vec{p}$ & N-body \& Hydrodynamic\\
\hline
$\Omega_{\rm m}$ & 17.4 \% & $V$  & - & N-body \& Hydrodynamic\\
\hline
$\Omega_{\rm m}$ & 5.6 \% & $V$ & $\vec{p}$ & N-body \& Hydrodynamic \\
\hline
$\sigma_{\rm 8}$ & 11.8 \% & - & $\vec{p}$ & N-body (halos $\gtrsim 300$ particles) \\
\hline
$\sigma_{\rm 8}$ & 5.9 \% & $M$ & $\vec{p}$  & N-body (halos $\gtrsim 300$ particles)\\
\hline
$\sigma_{\rm 8}$ and $\Omega_{\rm m}$ & 2.7, 2.2\% & $V$, $c$, $V_{\rm max}$ & $\vec{p}$ &  Not Robust \\
\hline
\end{tabular}
\end{center}
\caption{This table lists the mean relative error, robustness, and input node features for each model that we discuss in Section \ref{sec:results}.  
\label{tab:models}}
\end{table*}

Finally, we perform further tests to gauge the extent of the robustness of the model:
\begin{itemize}
    \item Robustness to resolution. We test the model on N-body simulations run at higher resolution --$512^3$ particles in a $(25~h^{-1}{\rm Mpc})^3$ volume--, finding that the model is still robust. We show the results in Fig. \ref{fig:high_resolution} of Appendix \ref{sec:HR}.
    \item Robustness to redshift. We evaluate the model, which was trained at redshift $z=0$, on halo catalogues generated at redshifts $z>0$. We find that the model is not able to infer the correct cosmology in this case, as expected, since clustering changes with redshift. We show the results in Fig. \ref{fig:z1} in Appendix \ref{sec:higher_redshift}. 
    \item Robustness to halo finder. We test the model, which was trained using halo catalogues generated with \textsc{Rockstar}, on catalogues generated with a different halo finder, namely \textsc{SUBFIND}. We find that the model is also not robust. Results are reported in Appendix \ref{sec:subfind}. Note that in \citet{calibration_of_hmf} authors studied in detail the abundance and clustering of halos from different halo finders, finding differences between Rockstar and SUBFIND, that could explain our results.
\end{itemize}

\subsection{Inferring \texorpdfstring{$\sigma_{\rm 8}$}{S8} \label{subsec:inferring_sigma8}}

We now focus our attention into the models trained to infer $\sigma_{\rm 8}$. When training using catalogues that only contain halo positions we obtain very weak constraints, with a relative error of $\epsilon = 11.8 \%$ and $R^2 = 0.27$. We have also trained a MLP to predict $\sigma_{\rm 8}$ from the halo power spectrum and found that the model is is unable to extract much information, performing with a mean relative error of $11.9\%$ and $R^2 = 0.28$, similar to the results from the GNN.
%The better results obtained using the GNNs show that there is much more information at the halo field level, similarly to what we found for $\Omega_{\rm m}$. 

In order to improve our constraint, we include additional halo properties. We find that using positions and velocities does not improve constraints significantly, so we train models using catalogues that contain halo positions and masses\footnote{Note that a model trained on catalogues that contain halo positions, velocities, and masses achieves a similar accuracy than the model trained on halo positions and masses.}. 

Our model performs well and is able to achieve a mean relative error of $\epsilon = 5.9 \%$ and a Chi squared value of $\chi ^2 = 1.54$. We show the results of testing the model on halo catalogues from Gadget in the right panel of Fig. \ref{fig:gadget_individual}. As it can be seen, the model is able to infer the correct value of $\sigma_8$ with a relatively high precision and accuracy.

\subsubsection{Robustness}

To investigate the robustness of the model we test it on halo catalogues from different N-body simulations.  In Fig. \ref{fig:same_seed} we show the results of testing the model on simulations run with different codes but sharing the same cosmology and initial random seed. We find that while the model can infer the correct value of $\sigma_8$ on all different halo catalogues from Abacus, CUBEP$^3$M, and PKDGrav3, it fails on halo catalogues from Enzo and Ramses that contain less than 300 particles. We reach similar conclusions when testing the models on all the other N-body simulations from the other codes that have different cosmologies and values of the initial random seed. Thus, from this test we conclude that the network is still robust to differences among N-body codes but requirements are more strict and halos with more particles (larger than 300) are needed in some codes to achieve robustness.

From Fig. \ref{fig:same_seed} we can also see that the models is not robust to hydrodynamics and astrophysical effects, as it fails to recover the true value of $\sigma_8$ from the halo catalogues of the hydrodynamic simulations. This indicates that the estimator used by the network is more affected by baryonic effects than the one used to infer $\Omega_{\rm m}$. There could be several reasons for this. For instance, this network employs the halo mass as an input node feature which can be heavily affected by galaxy formation feedback processes present in hydrodynamic simulations. Interestingly, the model becomes more accurate for more massive halos. We speculate that this could be explained with the consideration that baryonic effects might less severely affect the properties and/or abundance of more massive halos. 

Similar to $\Omega_{\rm m}$, we test the model on higher resolution N-body simulations and find that the predictions are robust across the different simulation codes. We report the results of this test in Fig. \ref{fig:high_resolution} of Appendix \ref{sec:HR}. We also test this model on catalogues at higher redshift ($z=1$) than the one used for training ($z=0$), finding that the model does not work in this case. We show the details in Appendix \ref{sec:higher_redshift}. Finally, we test the model with catalogues generated with a different halo finder. In this case that the model is not robust; we show the results in Appendix \ref{sec:subfind}.

\subsection{Dependence on halo properties}

We have seen that our models are robust to numerical differences across N-body codes when halo catalogues contain positions plus velocities and positions plus masses. We expect that adding more properties to the halo catalogues will make our models more precise, but will they be robust?

To test this we train a GNN using Gadget catalogues that contain halo positions, velocity moduli ($V$), concentrations ($c$), and maximum circular velocities ($V_{\rm max}$) to infer both $\Omega_{\rm m}$ and $\sigma_8$. We find that our model achieves a higher accuracy and is able to infer the value of $\Omega_{\rm m}$ and $\sigma_8$ simultaneously, when tested on Gadget catalogues, with a mean relative error of $2.7\%$ and $2.2\%$ respectively.

Unfortunately, the model is no longer robust, and when tested on halo catalogues from other simulations it fails. We show the results and further details on this in the Appendix \ref{sec:both}. We note that halo properties such as masses or velocities receive large contributions from the halo outskirts, where a large number of particles are. On the other hand, properties such as concentration and $V_{\rm max}$ weight more heavily the distribution of particles near the halo center. That region, well inside the 1-halo term, is where the differences among codes will appear more clearly. We believe that this is the reason for the failure of our model.

This exercise raises an important point that marks the line between accuracy and precision. While it is possible to make models more precise by considering more properties, the accuracy may be severely affected as the model will be learning more features and properties that are not shared across codes. Therefore, it is important to build models that are both precise and accurate.

\section{Conclusions} \label{sec:conclusion}

In this work we have addressed the question of how much robust information can be extracted from halo catalogues. This question was motivated by \citet{pablo}, who found that their models were not robust when using galaxy catalogues from hydrodynamic simulations. Since the optimal estimator to extract cosmological information from non-linear scales is unknown, we have trained graph neural networks to perform field-level likelihood-free inference. As already outlined in \cite{pablo}, the advantage of working with GNNs is that they are, by construction, permutationally invariant and are designed to handle sparse and irregular data without impose a minimum scale. Furthermore, we build our models to fulfill translationally and rotationally symmetries.

The input to our models are thousands of halo catalogues generated from Gadget N-body simulations. Each catalogue contains less than $\sim5,000$ halos in a periodic volume of $(25~h^{-1}{\rm Mpc})^3$ that are characterized by several properties such as position, mass, velocity, concentration, and maximum circular velocity. We summarize the main takeaways of this work below:

\begin{itemize}
    
    \item We find that our field-level inference using catalogues that only have halo positions achieve much tighter constraints than the ones obtained using the traditional power spectrum. 
    
    \item The GNNs trained using catalogues that contain the halo positions and velocity moduli are able to infer the value of $\Omega_{\rm m}$ with a mean relative error of 5.5\%. We find that most of the information is coming from the clustering of halos, not from the velocity distribution of the catalogue.
    
    \item The model is surprisingly robust, and is able to infer the value of $\Omega_{\rm m}$ from thousands of halo catalogues generated from five different N-body codes --Abacus, CUBEP$^3$M, Enzo, PKDGrav3, Ramses-- and four different hydrodynamic codes with different galaxy formation implementations --Astrid, IllustrisTNG, Magneticum, SIMBA--. We emphasize that this is not a trivial result since the different N-body codes solve the equations using different methods and approximations and the hydrodynamic simulations implement completely different recipes for astrophysical processes such as supernova and AGN feedback, that are expected to affect the abundance and properties of dark matter halos.
    
    \item The GNNs trained using catalogues that contain the halo positions and masses are able to infer the value of $\sigma_8$ with a mean relative error of $5.7\%$. The model is robust across different N-body simulations only for catalogues that contain halos with more than $\sim300$ particles. However, this model is not robust when tested on catalogues from hydrodynamic simulations. On the other hand, the model becomes more accurate for catalogues with more massive halos. We thus conclude that the requirements for robust inference of $\sigma_8$ are more stringent than for $\Omega_{\rm m}$.

    \item The GNNs trained using catalogues that contain halo positions, velocity moduli, concentrations, and maximum circular velocities ($V_{\rm max}$) are able to extract more information, and they can infer the value of $\Omega_{\rm m}$ and $\sigma_{\rm 8}$ with mean relative errors of $\epsilon\sim2.7 \%$ and $\epsilon\sim2.2 \%$, respectively. However, the use of these properties decreases the robustness of the model: the model is not longer robust across N-body codes. 
    
\end{itemize}

The surprising robustness of the model trained to infer $\Omega_{\rm m}$ may be due to the existence of an underlying fundamental relation between halo positions and velocities and $\Omega_{\rm m}$ that is not largely affected by numerical errors from different N-body codes or baryonic effects. In future work we plan to use interpretation tools to try to understand how the network is performing the inference.

It has to be emphasized the difficulties of obtaining robust models across different N-body and hydrodynamic codes. For instance, already with summary statistics different codes exhibits different levels of discrepancies \citep{Heitmann_2008, schneider, 2019MNRAS.485.3370G}. Nevertheless, having robust predictors is essential in order to apply them to observational data. Only models which are proven to be robust over different simulation methods could offer trustworthy results when tested on observational catalogues.

In this paper we have studied the regime where halo catalogues will enable the construction of robust models. However, dark matter halos are not directly observable, but galaxies are. \citet{pablo} studied how much information can be extracted from galaxy catalogues through GNNs, but found that the models are not robust. However, they trained their models using both centrals and satellite galaxies. In future work we plan to investigate whether the constraints from galaxy catalogues will be more robust if only central galaxies are included in the catalogues. The idea behind this is be that the positions and velocities of the central galaxies should be similar to those of the halo they reside in.

We note that all the analysis performed in this work has been carried out in real-space. In future work we plan to repeat this analysis but using halos in redshift-space and investigate whether the accuracy and robustness of the models are affected by considering redshift-space distortions.

In this work we have used relatively small cosmological boxes of $(25~h^{-1}{\rm Mpc})^3$ volume. The reason for this was because we could take advantage of the thousands of N-body and hydrodynamic simulations publicly available from the CAMELS project. While it would have been desirable to use large boxes that sample more massive halos, we could not carried out the same analysis we have done here with hydrodynamic simulations. On the other hand, given the small volume of our simulations, the effects of super-sample covariance can be pretty important. It would be important to redo the analysis performed in this work but with larger boxes, even if limited to N-body simulations, to investigate the impact of large-scale modes on our results. We note that training our models on larger volume catalogues, at similar number density, will naturally increase the precision of the models. Thus, it will be important to quantify whether the models still remain accurate. 

Finally, this work emphasizes the trade-off between accuracy and robustness. While more information can be extracted from catalogues that contain halo properties such as concentration and $V_{\rm max}$, that information is not robust, and different N-body codes produce slightly different answers that break down the robustness of the prediction. While using the information from those properties should not be disregarded, its inclusion in the model may require higher demands; e.g. only considering halos with more than 1,000 particles where all codes yield similar predictions. 

On the other hand, the use of properties that are very sensitive to the code used may never yield robust constraints. For instance, information from scales close to the softening length may always be very different in different N-body codes.

This work represent the first attempt to build robust models when performing field-level likelihood-free inference. Our goal is to develop models that work across galaxy catalogues from different hydrodynamic simulations, semi-analytic models, and other methods (e.g. Halo Occupation Distribution).

\section{Acknowledgments}

We thank Douglas Potter for his help with the PKDGrav3 simulations. We also thank Tom Abel, Simeon Bird, Shy Genel, and David Spergel for enlightening discussions. The networks have been trained using the Tiger cluster at Princeton University. HS thanks the Flatiron Institute for the support during the preparation of this work. The work of FVN has been supported by NSF grant AST-2108078. EV is supported by NSF grant AST-2009309 and NASA grant 80NSSC22K0629. DAA acknowledges support by NSF grants AST-2009687 and AST-2108944, CXO grant TM2-23006X, and Simons Foundation award CCA-1018464. TC is supported by the INFN INDARK PD51 grant and by the FARE MIUR grant `ClustersXEuclid' R165SBKTMA. EHM was supported by the grant agreements ANR-21-CE31-0019 / 490702358 from  the French Agence Nationale de la Recherche / DFG for the LOCALIZATION project. KD acknowledges support through the COMPLEX project from the European Research Council (ERC) under the European Union’s Horizon 2020 research and innovation program grant agreement ERC-2019-AdG 882679 as well as support by the Deutsche Forschungsgemeinschaft (DFG, German Research Foundation) under Germany’s Excellence Strategy - EXC-2094 - 390783311. The CAMELS project is supported by NSF grants AST-2108944, AST-2108678, and AST-21080784. The Flatiron Institute is supported by the Simons Foundation. Kavli IPMU is supported by World Premier International Research Center Initiative (WPI), MEXT, Japan.

\appendix

\section{Epistemic Errors}\label{sec:epistemic}

\begin{figure*}[h]
    \centering
    \includegraphics[width=1\textwidth]{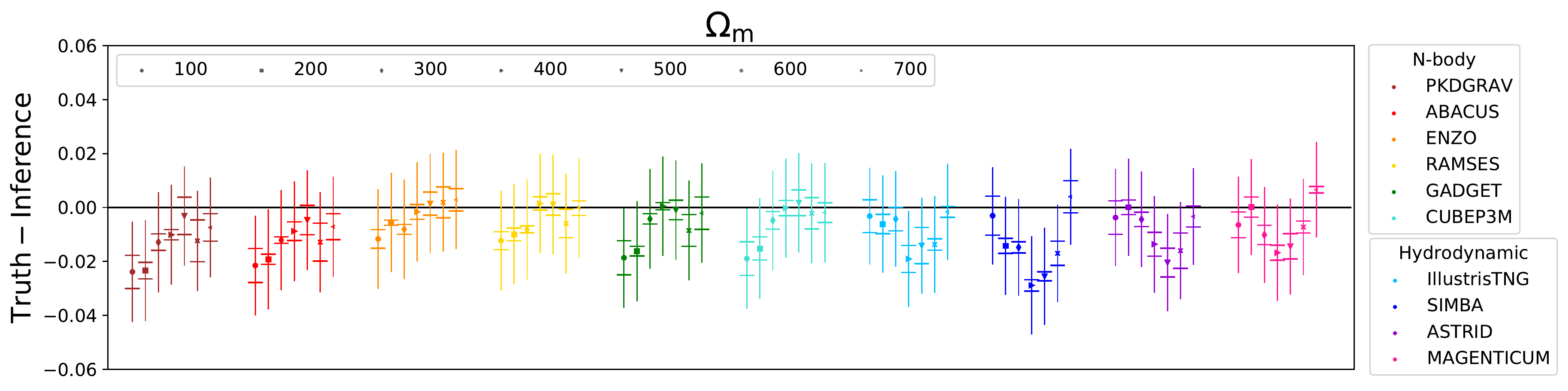}
    \includegraphics[width=1\textwidth]{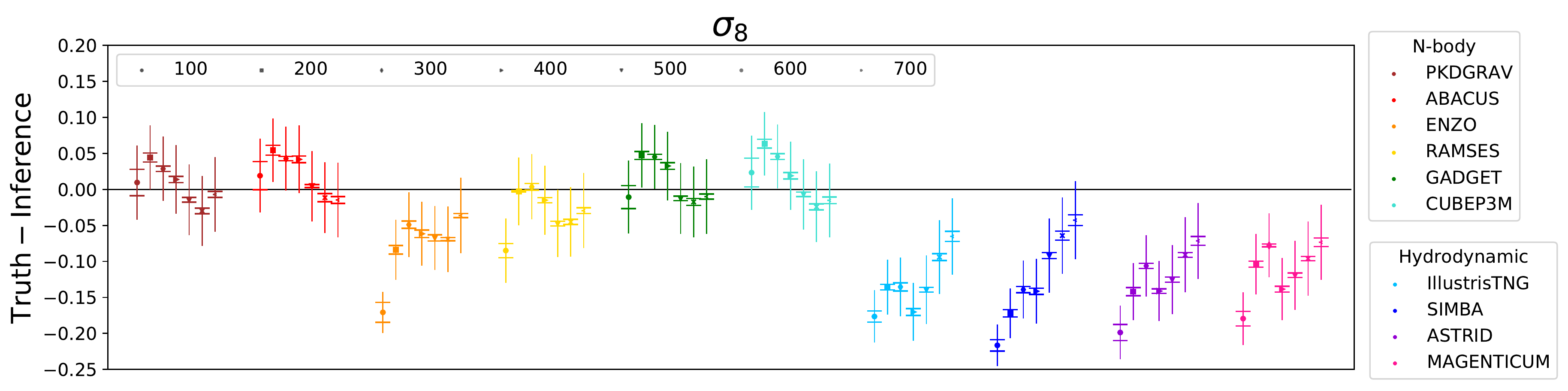}
    \caption{We compute the epistemic errors for each halo catalogue, denoted by the horizontal error bar caps centered around the epistemic means, for the models trained to infer $\Omega_{\rm m}$ and $\sigma_{\rm 8}$. The vertical error bars denote the aleatoric errors which are the predicted standard deviations. As it can be seen, the epistemic errors are small compared to the aleatoric errors which indicate that the trained models do not exhibit large variability in their weights.}
    \label{fig:epistemic}
\end{figure*}

Here, we report the epistemic uncertainties associated with the models trained to predict $\Omega_{\rm m}$ and $\sigma_{\rm 8}$, which quantify the variability in the weights of the neural networks. To calculate these errors, we train five GNNs to predict the respective cosmological parameter with different initializations of the network weights. For this, we fix the hyperparameter values that were already optimized, including the linking radius, the number of hidden layers, the number of hidden features, the learning rate, and the weight decay. We also employ the same training and validation datasets across the five models. We then test each of the five models on each halo catalogue for the different N-body and hydrodynamic simulations. We compute the epistemic errors by taking the standard deviations of the predictions for each halo catalogue. We find that they are small, averaging to be $0.4\%$ and $0.6\%$ for the $\Omega_{\rm m}$ and $\sigma_{\rm 8}$ models, respectively, across the different particle thresholds of all simulations. Note that the aleatoric errors we estimate are $\sim6\%$ for both $\Omega_{\rm m}$ and $\sigma_8$. In Fig. \ref{fig:epistemic} we illustrate the epistemic errors with errorbar caps for each halo catalogue centered around the epistemic means. We thus conclude that that error budget is dominated by the aleatoric error, rather than the epistemic one.

\section{Additional Plots on \texorpdfstring{$\Omega_{\rm m}$}{Om} and \texorpdfstring{$\sigma_{\rm 8}$}{S8}\label{sec:additional plots}}

\begin{figure*}
    \centering
    \includegraphics[width=.45\textwidth]{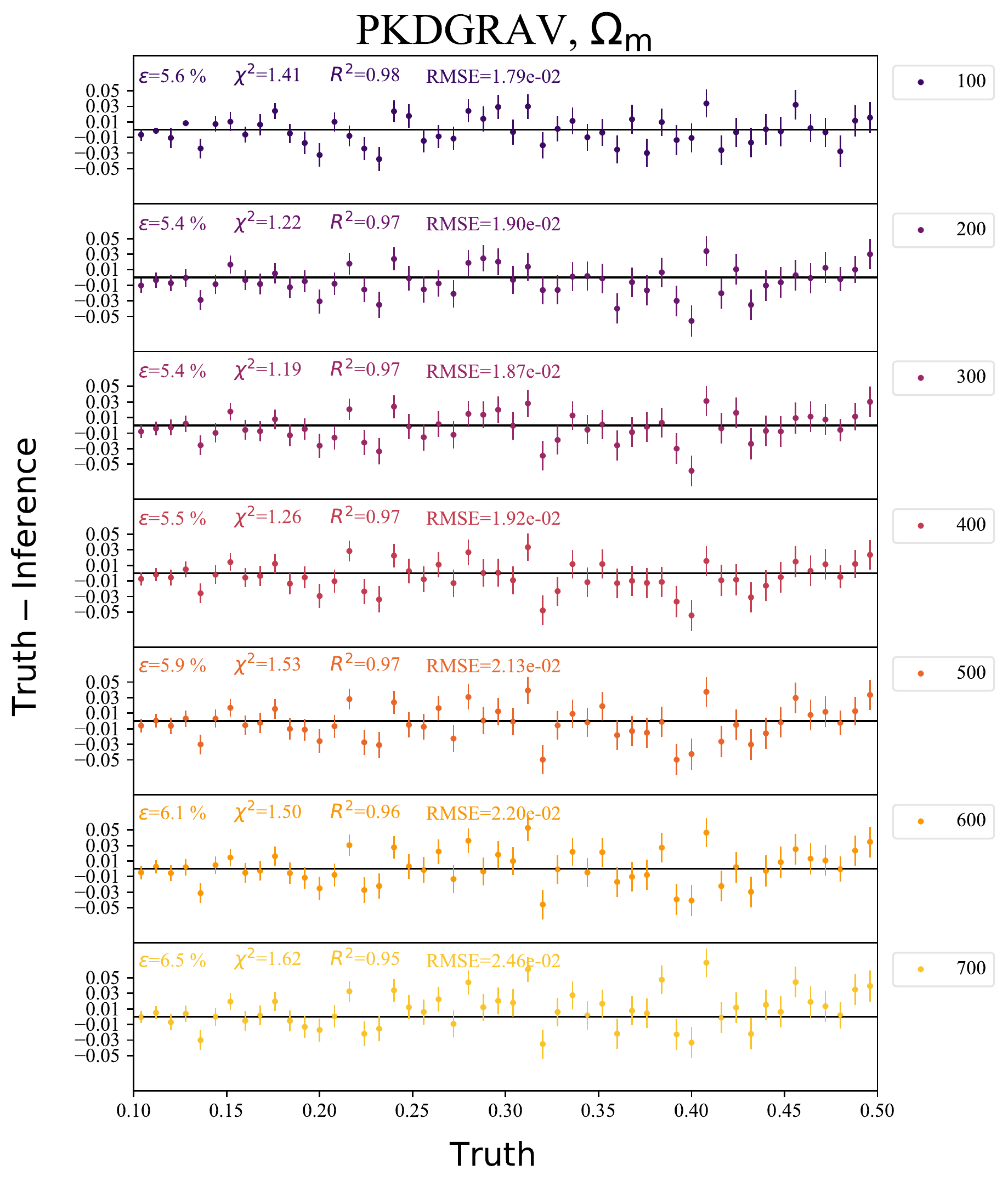}
    \includegraphics[width=.45\textwidth]{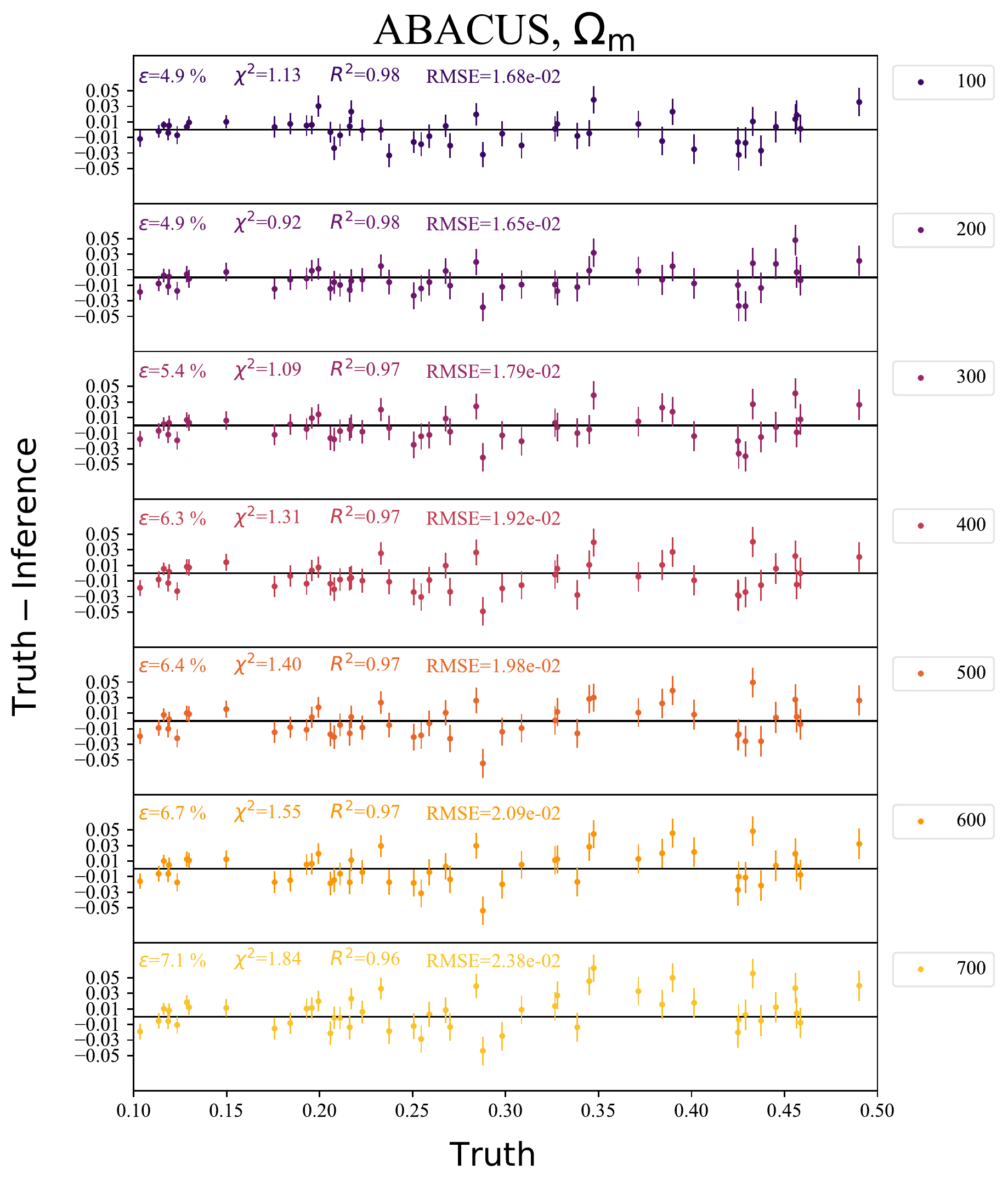}
    \includegraphics[width=.45\textwidth]{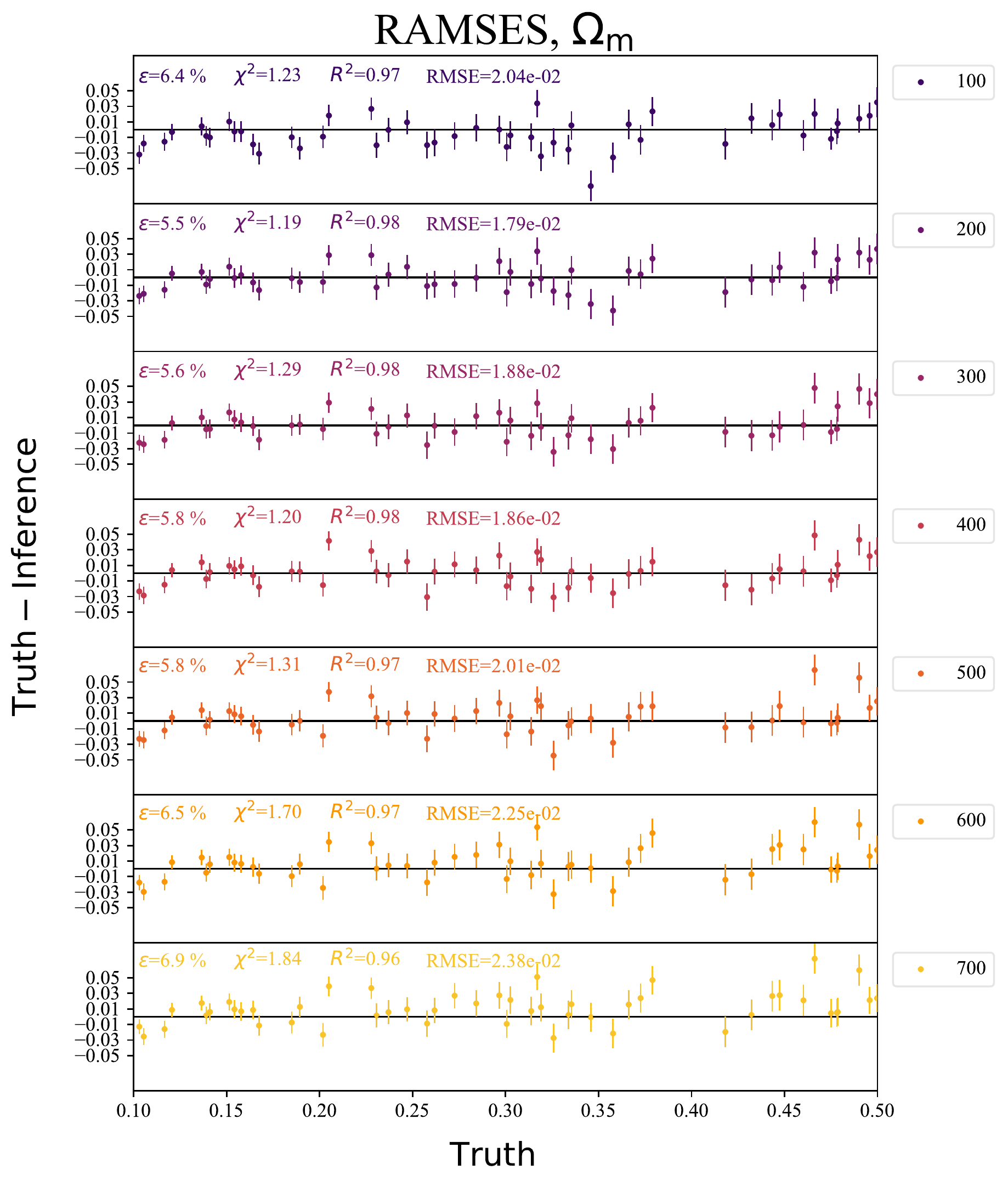}
    \includegraphics[width=.45\textwidth]{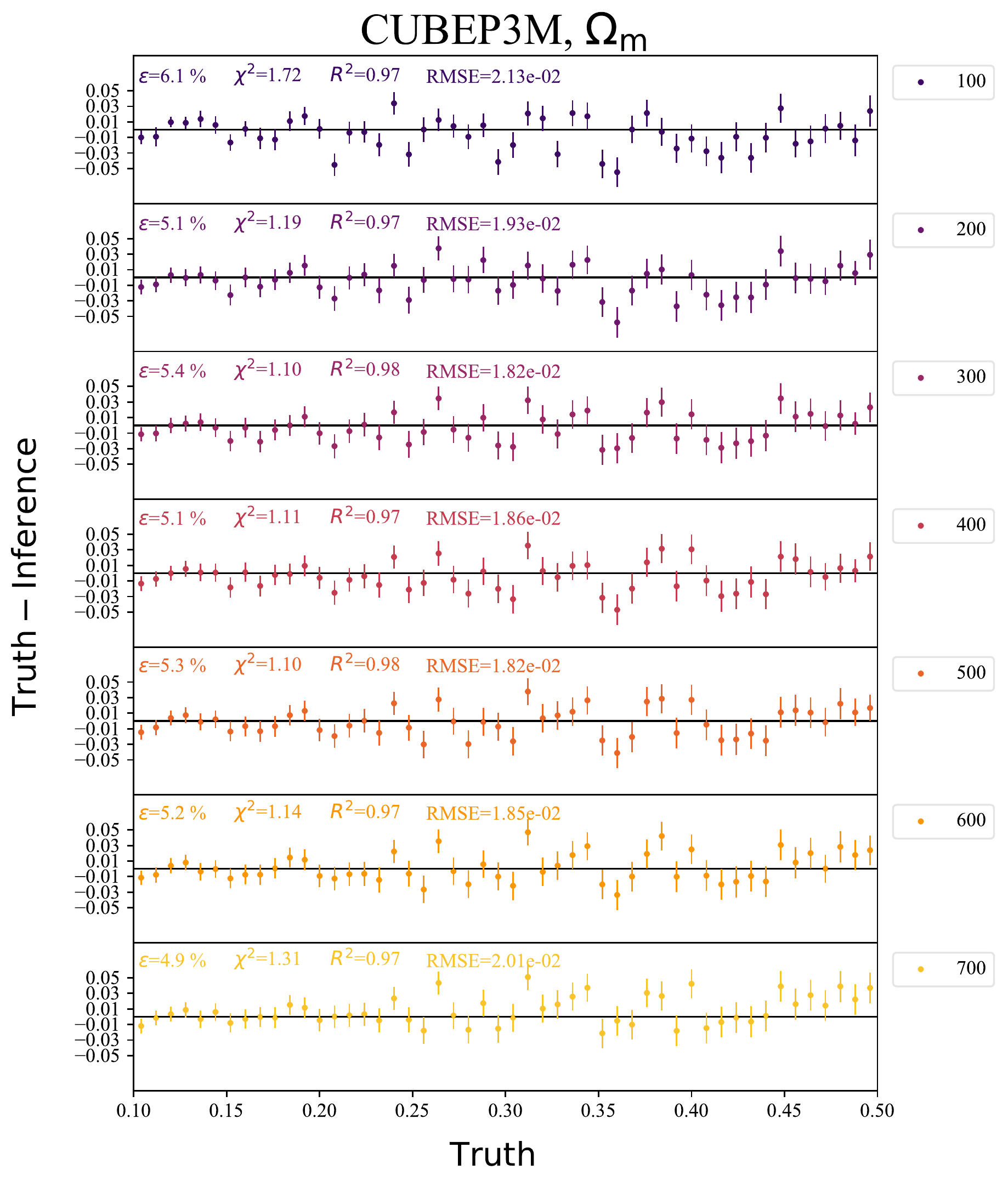}
    \caption{We train a GNN to infer $\Omega_{\rm m}$ from catalogues of the Gadget N-body simulations using the halo relative positions and velocity moduli. We teach the model to marginalize over the number density of halos in each catalogue by training it using halo catalogues constructed with random minimum particle thresholds ranging from [100, 700]. We then test this model on different N-body simulations: PKDGrav3, Abacus, Ramses, and CUBEP$^3$M using catalogues created with particle thresholds indicated next to the plots. As can be seen, the model is able to extrapolate well to different N-body codes and is able to predict with high accuracy for halos of all masses, suggesting that the model has found a universal relation to predict the cosmological parameter.}
    \label{fig:omegaM_NBody}
\end{figure*}

\begin{figure*}
    \centering
    \includegraphics[width=.45\textwidth]{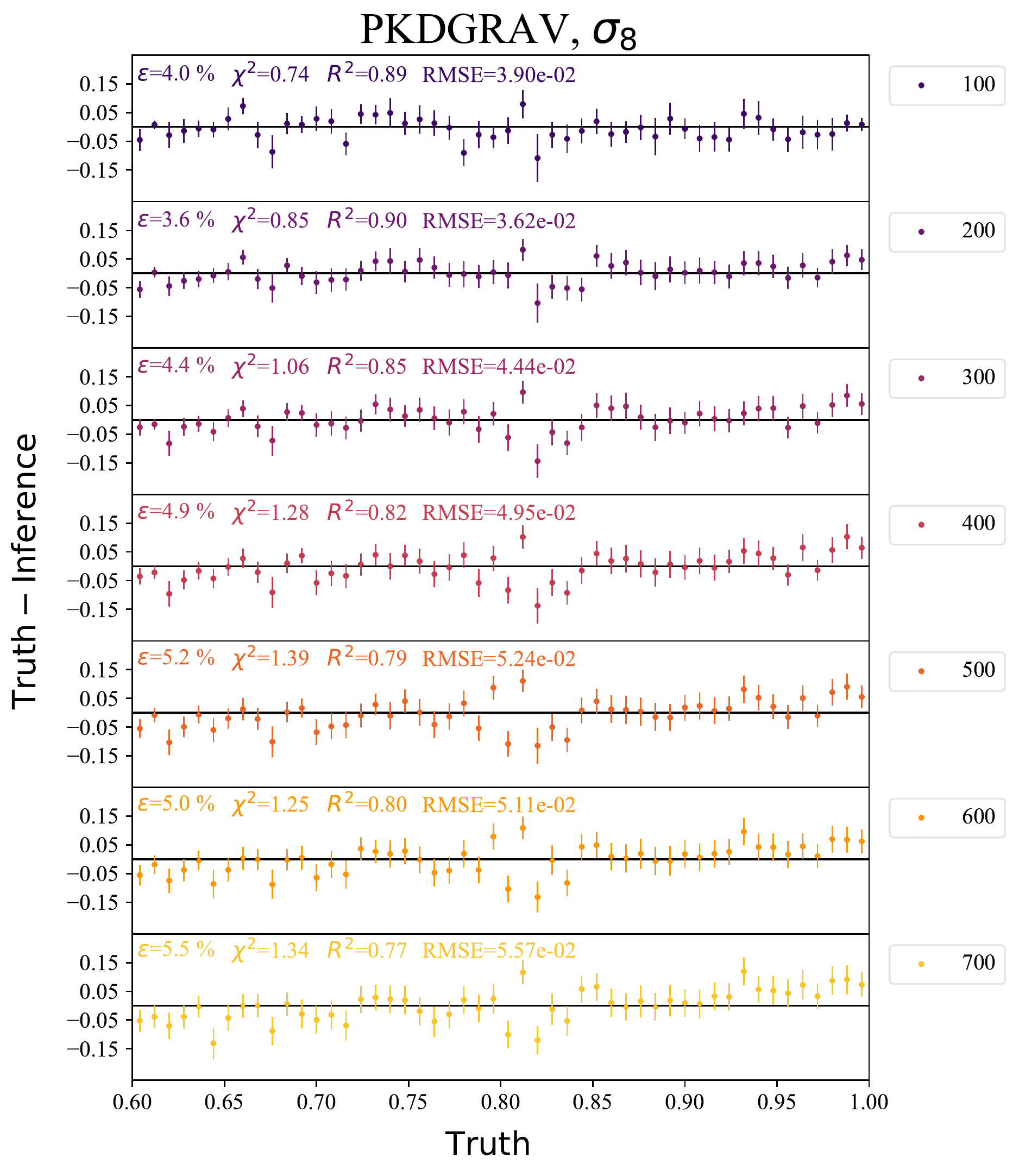}
    \includegraphics[width=.45\textwidth]{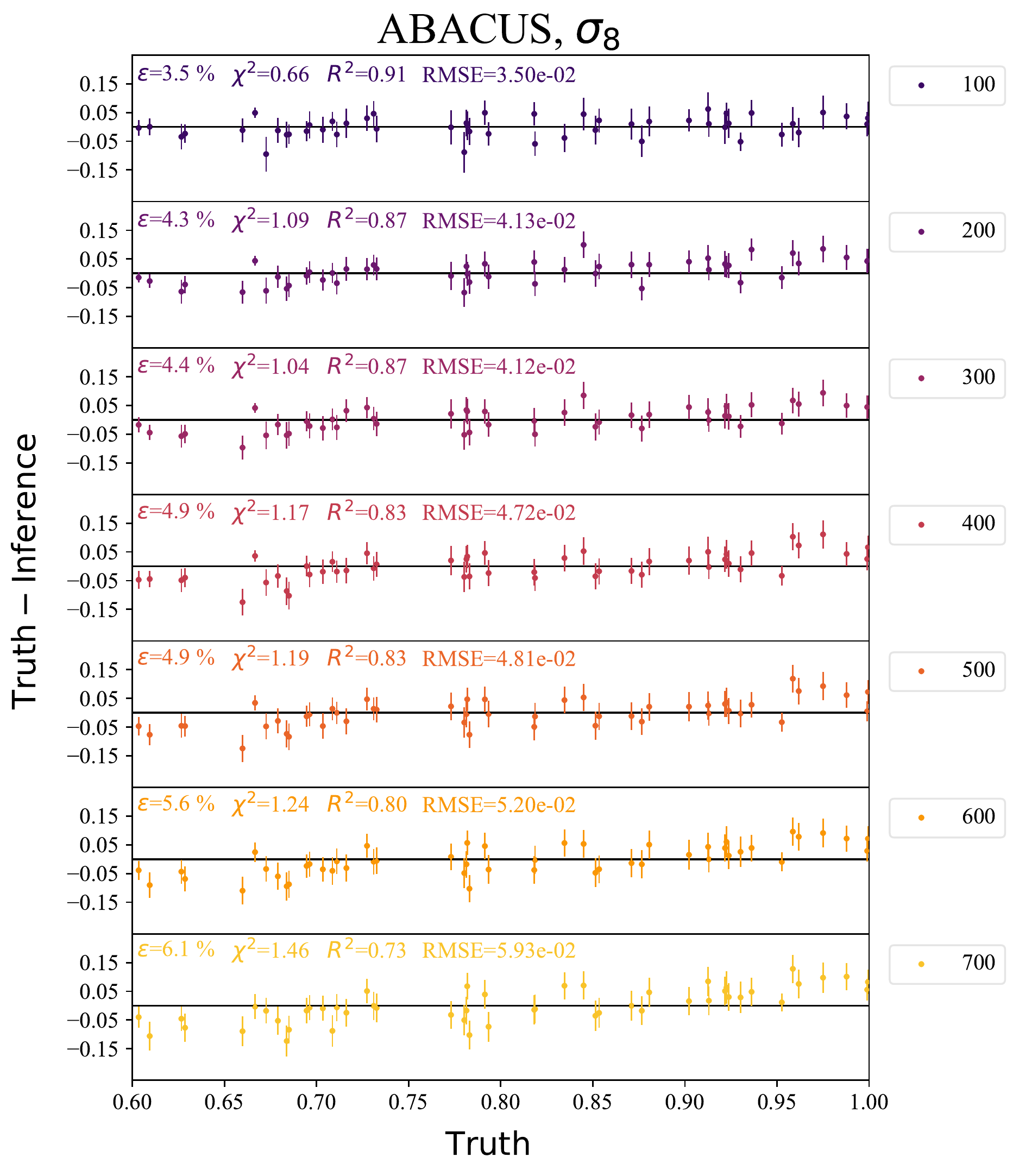}
    \includegraphics[width=.45\textwidth]{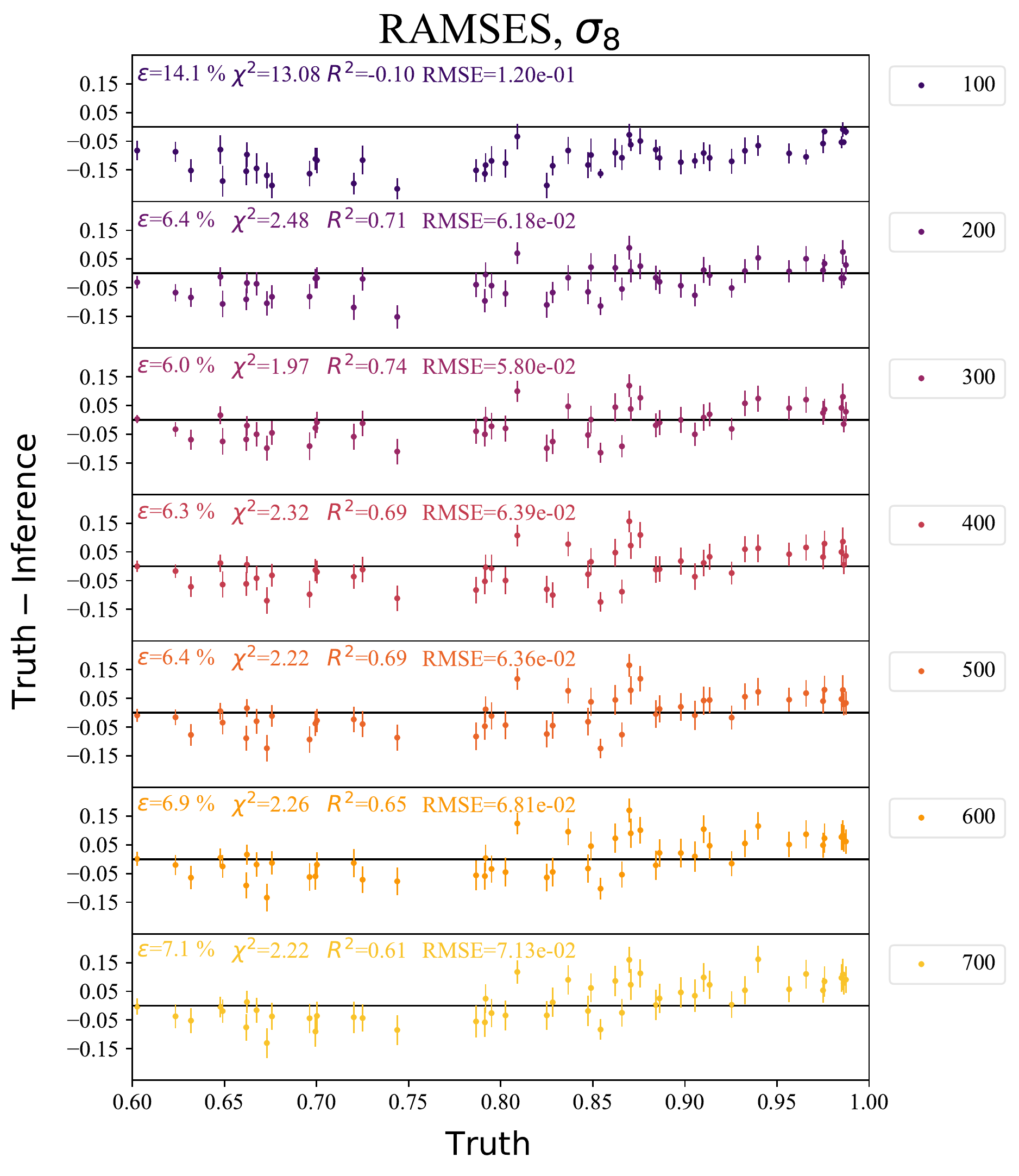}
    \includegraphics[width=.45\textwidth]{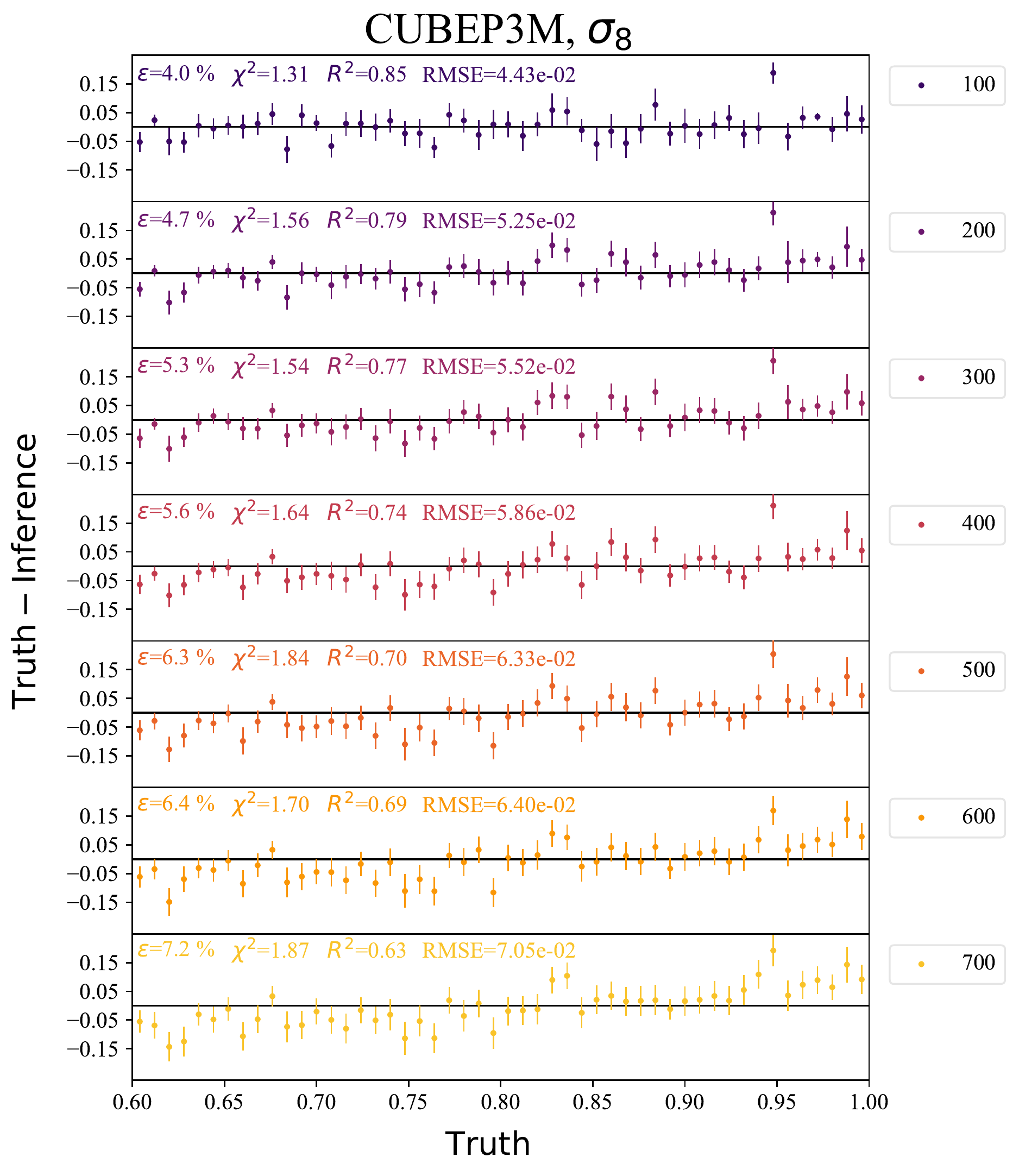}
    \caption{This is the same as Fig. \ref{fig:omegaM_NBody} for the parameter $\sigma_{\rm 8}$. To train this GNN, we use the halo relative positions and masses, and train with halo catalogues created from random minimum particle thresholds ranging from [100, 700]. Evidently, the model is not robust across different N-body codes for catalogues generated with low particle thresholds, as indicated by the significant bias in the predictions for Ramses catalogues created with particle thresholds of 100 or 200. However, the model is robust for larger particle thresholds, such as for halos that contain at least $\sim300$ particles, where the coefficients of determination are closer to 1 and the relative errors are $\sim 6\%$. Note that for all simulations, the slight increase in the mean relative errors and biases in the predictions for the catalogues created with particle thresholds of 600 or 700 particles is due to the fact these catalogues contain significantly fewer halos which hinders the model training, and should not suggest that the model fails to be robust across simulation codes for these massive halos.}
    \label{fig:sigma8_NBody}
\end{figure*}

\begin{figure*}
    \centering
    \includegraphics[width=.45\textwidth]{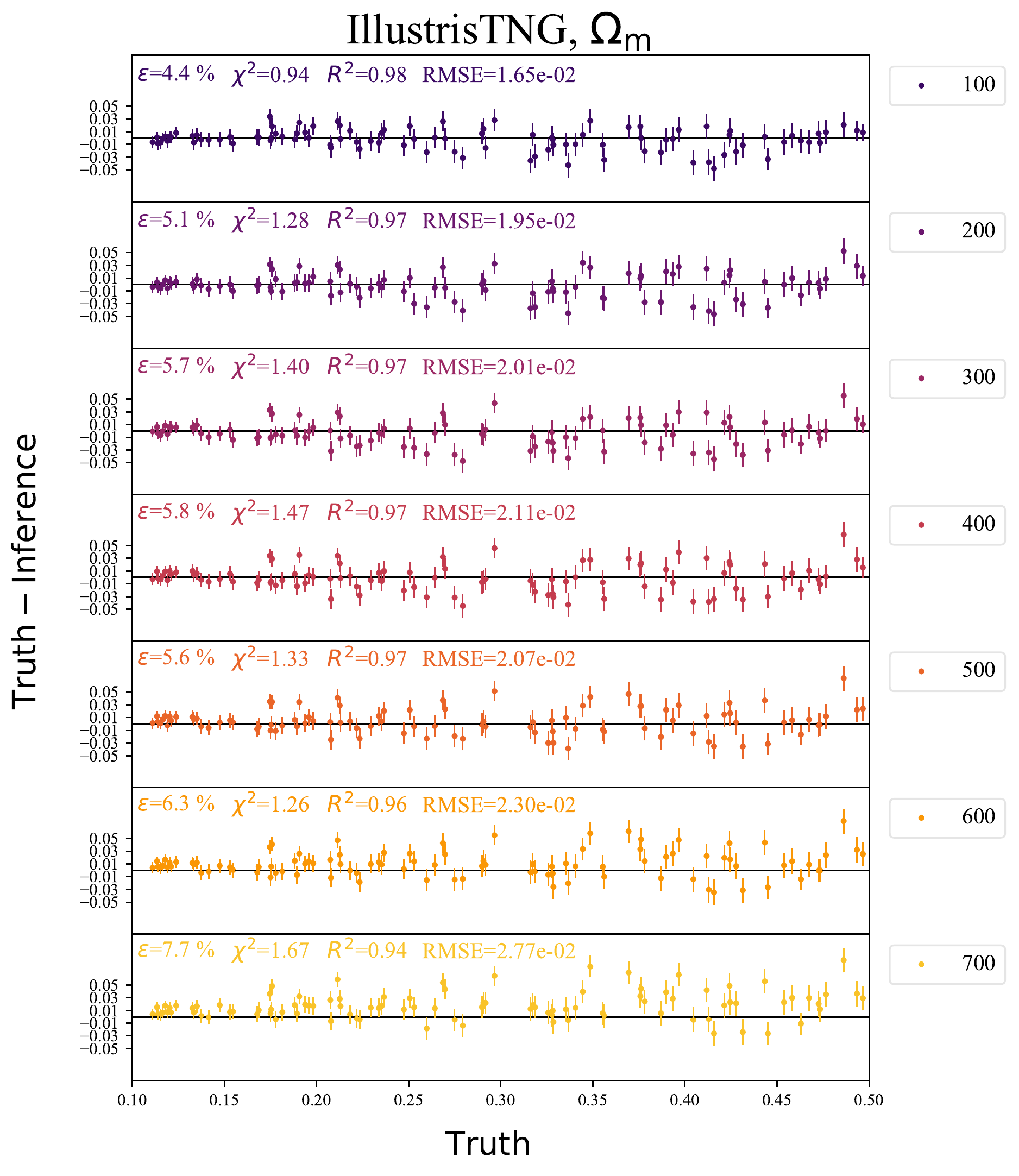}
    \includegraphics[width=.45\textwidth]{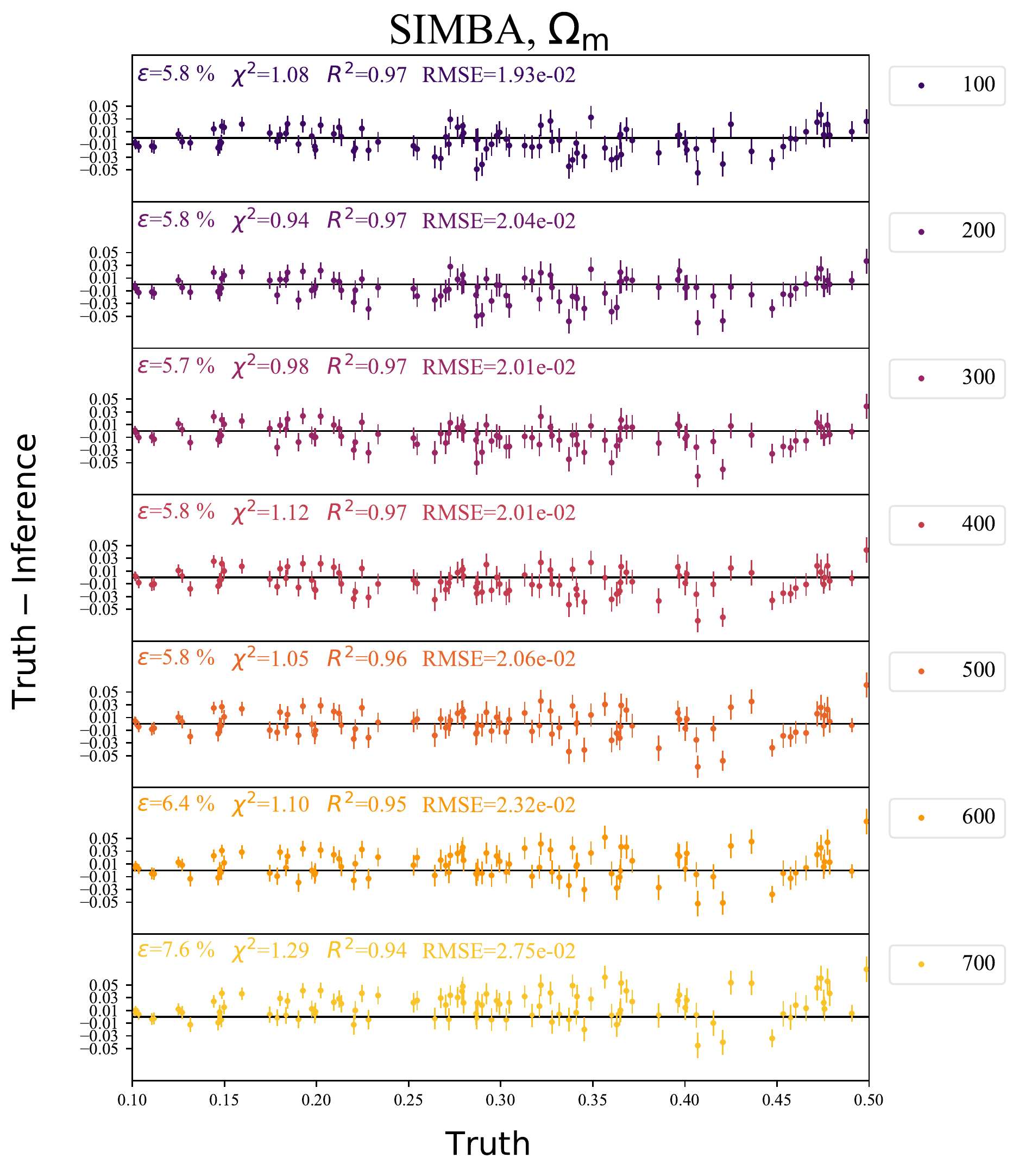}
    \includegraphics[width=.45\textwidth]{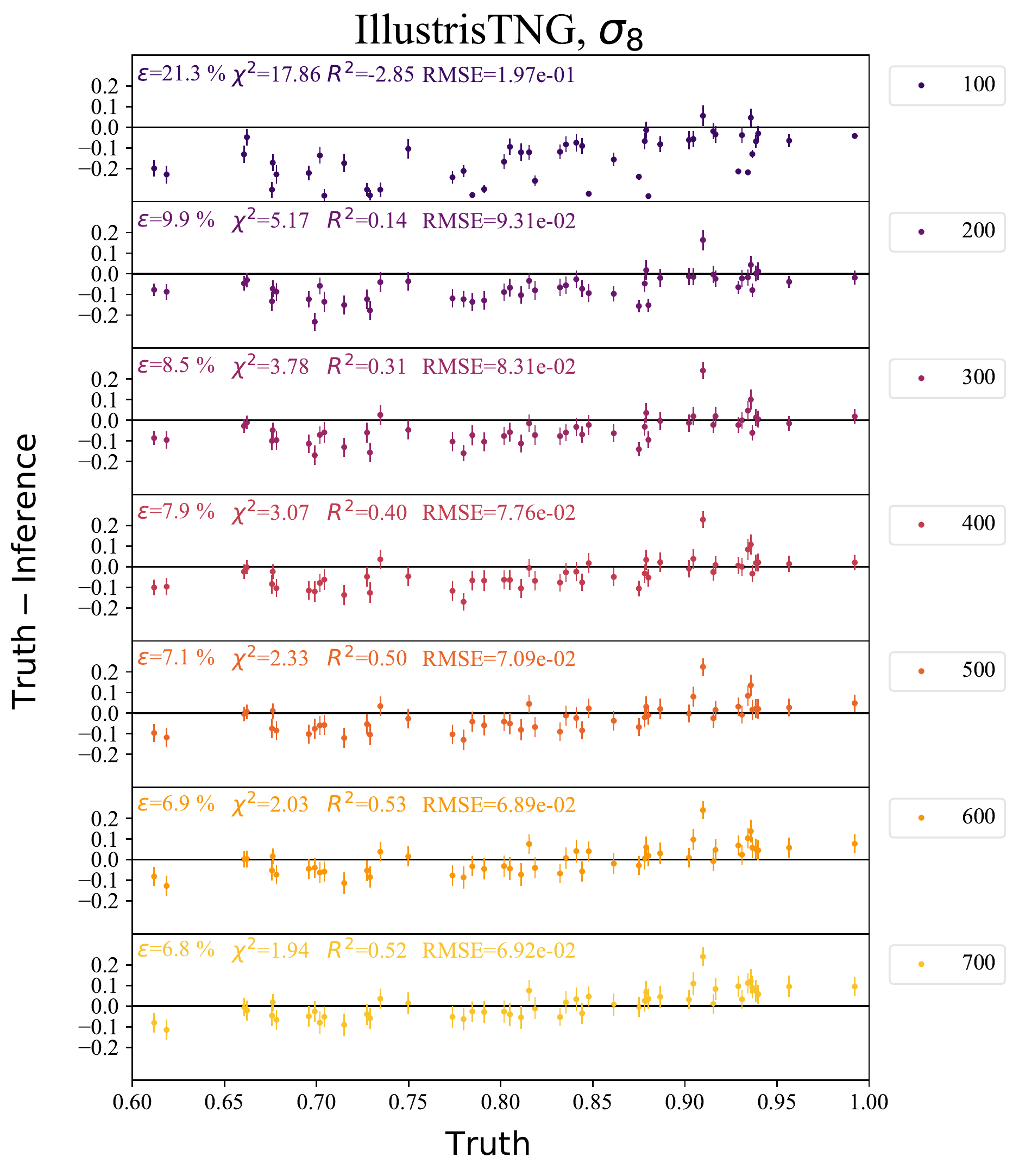}
    \includegraphics[width=.45\textwidth]{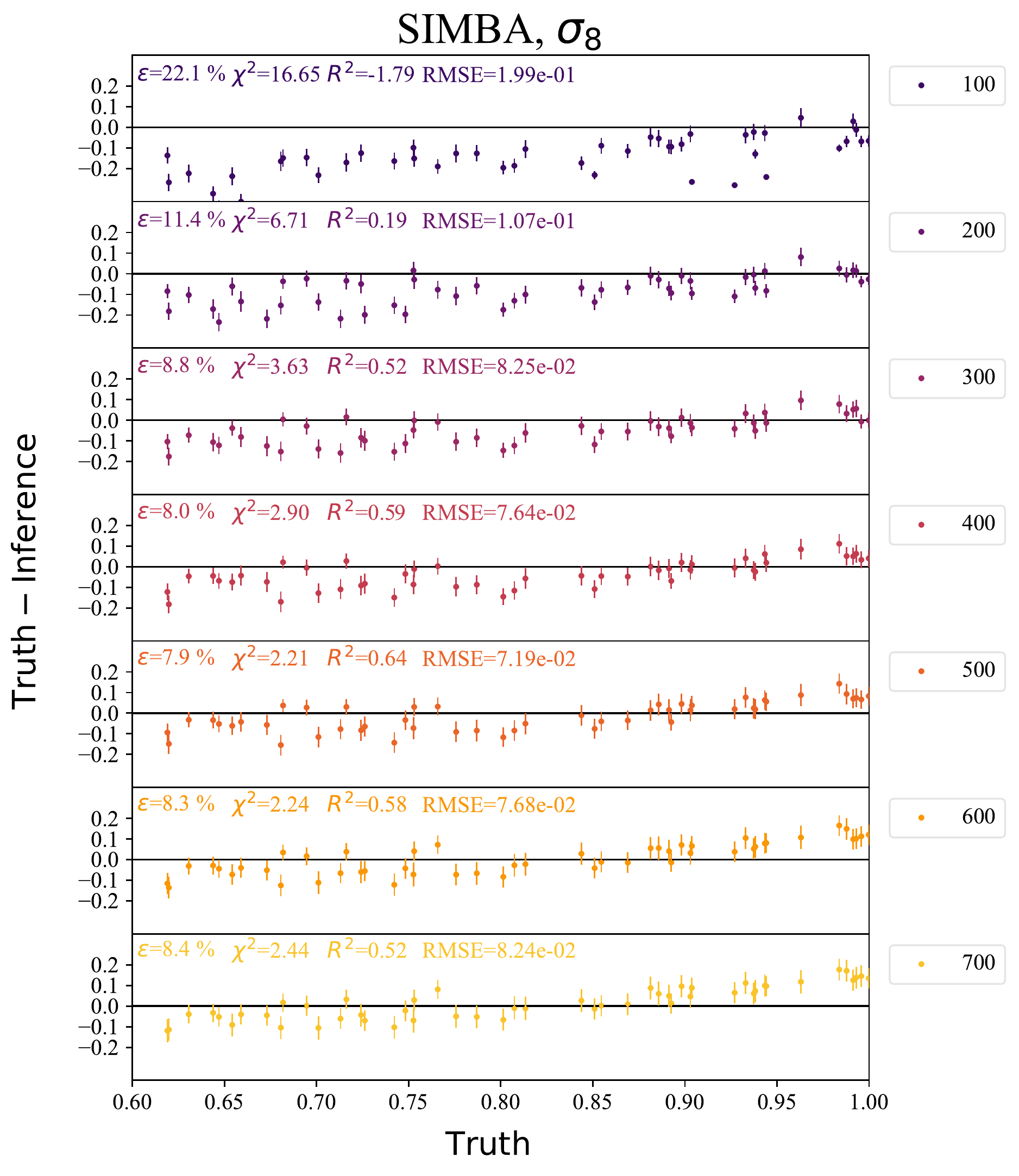}
    \caption{Similar to Figures \ref{fig:omegaM_NBody} and \ref{fig:sigma8_NBody}, we test the two models trained to infer $\Omega_{\rm m}$ and $\sigma_{\rm 8}$, respectively, on 1000 hydrodynamic simulations from IllustrisTNG and SIMBA that are run with different cosmological and astrophysical parameters. To conserve space, these plots show the results for 100 randomly selected simulations. As it can be seen, while the model that learned $\Omega_{\rm m}$ extrapolates well to the hydrodynamic simulations, the model that learned $\sigma_{\rm 8}$ does not. This can be seen in the plots for $\sigma_{\rm 8}$ where the predictions exhibit significant offsets from the truth. While these offsets are mitigated in catalogues of larger halos, such as those that contain more than 700 particles, the errors in the predictions still remain large and contain noticeable biases. This may be because $\sigma_{\rm 8}$ is more sensitive to the additional baryonic and astrophysical effects present in these simulations.}
    \label{fig:omegaM_sigma8_Hydro}
\end{figure*}

In Section \ref{sec:results}, we reported the results of our GNNs and their performance on various N-body and hydrodynamic simulations that were run with the same cosmologies and initial conditions. Here, we present additional plots depicting the accuracy of the model predictions for both $\Omega_{\rm m}$ and $\sigma_{\rm 8}$ for different minimum halo particle thresholds. For these plots, we evaluate the models on 50 simulations containing different cosmologies and initial conditions for four different N-body codes: Abacus, CUBEP$^3$M, PKDGrav, and Ramses (in Fig. \ref{fig:omegaM_NBody} and \ref{fig:sigma8_NBody}). We also test the models on 1,000 simulations from two hydrodynamic codes: IllustrisTNG and SIMBA, but plot the results for 100 randomly selected simulations to conserve space (in Fig. \ref{fig:omegaM_sigma8_Hydro}). As before, we perform these tests using halo catalogues created with different minimum halo particle thresholds as indicated in the plots. Each of these plots depict the predictions plotted against the truth minus the inference. 

As it can be seen, the GNN is able to infer $\Omega_{\rm m}$ accurately for all simulations with similar mean relative errors of $\sim 6\%$. However, the GNN trained to predict $\sigma_{\rm 8}$ is not robust to the hydrodynamic codes and is only robust across the different N-body simulations at high particle thresholds. This can be clearly observed in the bottom left panel of Fig. \ref{fig:sigma8_NBody} where the predictions exhibit a large offset from the truth for the lower thresholds ($\sim 100-200$ particles). Moreover, note that for the $\sigma_{\rm 8}$ predictions on N-body codes, there exists a slight increase in the mean relative errors for the catalogues created with thresholds of 600 or 700 particles due to larger biases in the predictions at the low and high ends of $\sigma_{\rm 8}$ values. This is due to the fact that these catalogues contain significantly fewer halos which hinders the model learning during training and should not suggest that the model fails to be robust across different simulation codes for these massive halos. A similar trend is seen in the predictions of $\Omega_{\rm m}$ with the hydrodynamic simulations. We believe that training with large simulation volumes that contain higher number densities of halos with at least 600 or 700 particles would improve the accuracies of the model predictions.

\section{Robustness to simulation resolution}\label{sec:HR}
\begin{figure*}
    \centering
    \includegraphics[width=1\textwidth]{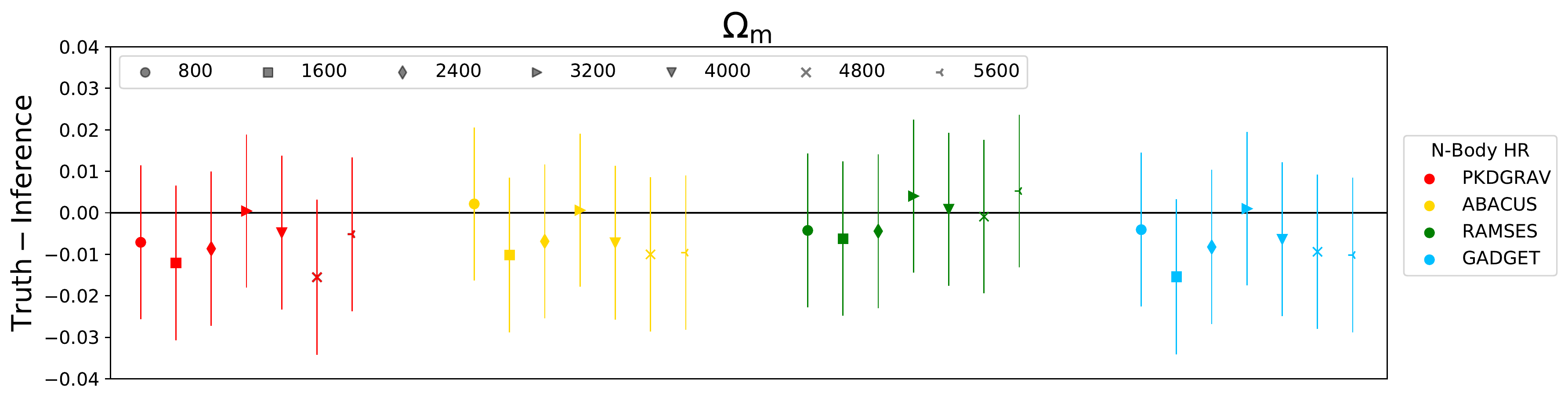}
    \includegraphics[width=1\textwidth]{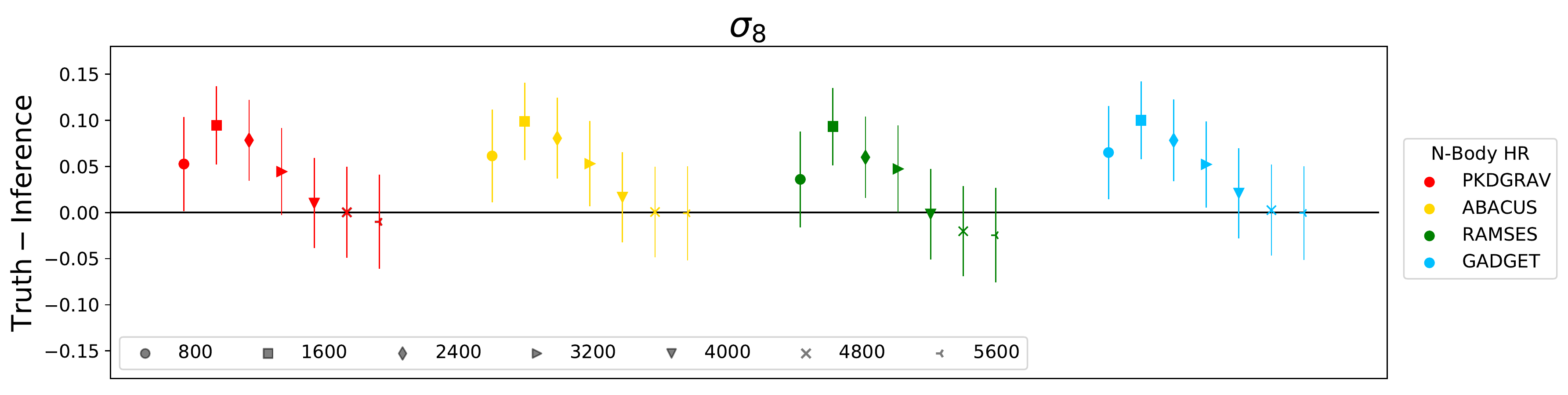}
    \caption{\textbf{Top:} We train a GNN to infer $\Omega_{\rm m}$ using Gadget N-body simulations with resolutions of $256^3$ particles and test it on different N-body simulations with resolutions of $512^3$ particles. The model is able to extrapolate well and performs with similar accuracy seen in Fig. \ref{fig:same_seed}. We note that the tests depicted here are evaluated on halo catalogues with minimum particle thresholds that are eight times larger than the ones used for the low resolution simulations due to the increase in number of particles, and consequently, the number of low-mass halos. \textbf{Bottom:} This is the same plot as above for testing the GNN trained to infer $\sigma_{\rm 8}$. As can be seen, the model is robust for halos with at least $3,200$ particles which agrees with the previously found results seen in Fig. \ref{fig:same_seed}.}
    \label{fig:high_resolution}
\end{figure*}

In Section \ref{sec:results} we trained two GNNs using Gadget N-body simulations that contain a resolution of $256^3$ particles to infer $\Omega_{\rm m}$ and $\sigma_{\rm 8}$. Here, we investigate whether these models are able to extrapolate to higher resolution simulations that contain $512^3$ particles. For this test, we use N-body simulations from PKDGrav3, Abacus, Ramses, and Gadget. It is important to note that due to the eight-fold increase in the number of particles, the tests are carried on halo catalogues defined with minimum particle thresholds that are eight times larger than the thresholds used for the lower resolution simulations. This is because higher resolution simulations contain more low-mass halos and this ensures that the model is being tested on halo mass ranges comparable to catalogues of the low resolution simulations. As it can be seen in the top panel of Fig. \ref{fig:high_resolution} the model is robust across the different higher resolution simulations for all minimum particle thresholds and is able to predict $\Omega_{\rm m}$ with comparable accuracy to the simulations with the resolution used for training. This further demonstrates that this model is utilizing a fundamental relation in the halo clustering that is present across all simulations. Similarly, the results for $\sigma_{\rm 8}$ are shown in the bottom plot of Fig. \ref{fig:high_resolution}. As evidenced, this model is also able to predict with high accuracy on the higher resolution simulations for halos that contain at least $3,200$ particles, corresponding to low resolution halos that contain at least $\sim300$ particles. This agrees with the previous findings that field-level inference for the parameter $\sigma_{\rm 8}$ is only robust for massive halos.

\section{Robustness to redshift}\label{sec:higher_redshift}

\begin{figure*}[h]
    \centering
    \includegraphics[width=.45\textwidth]{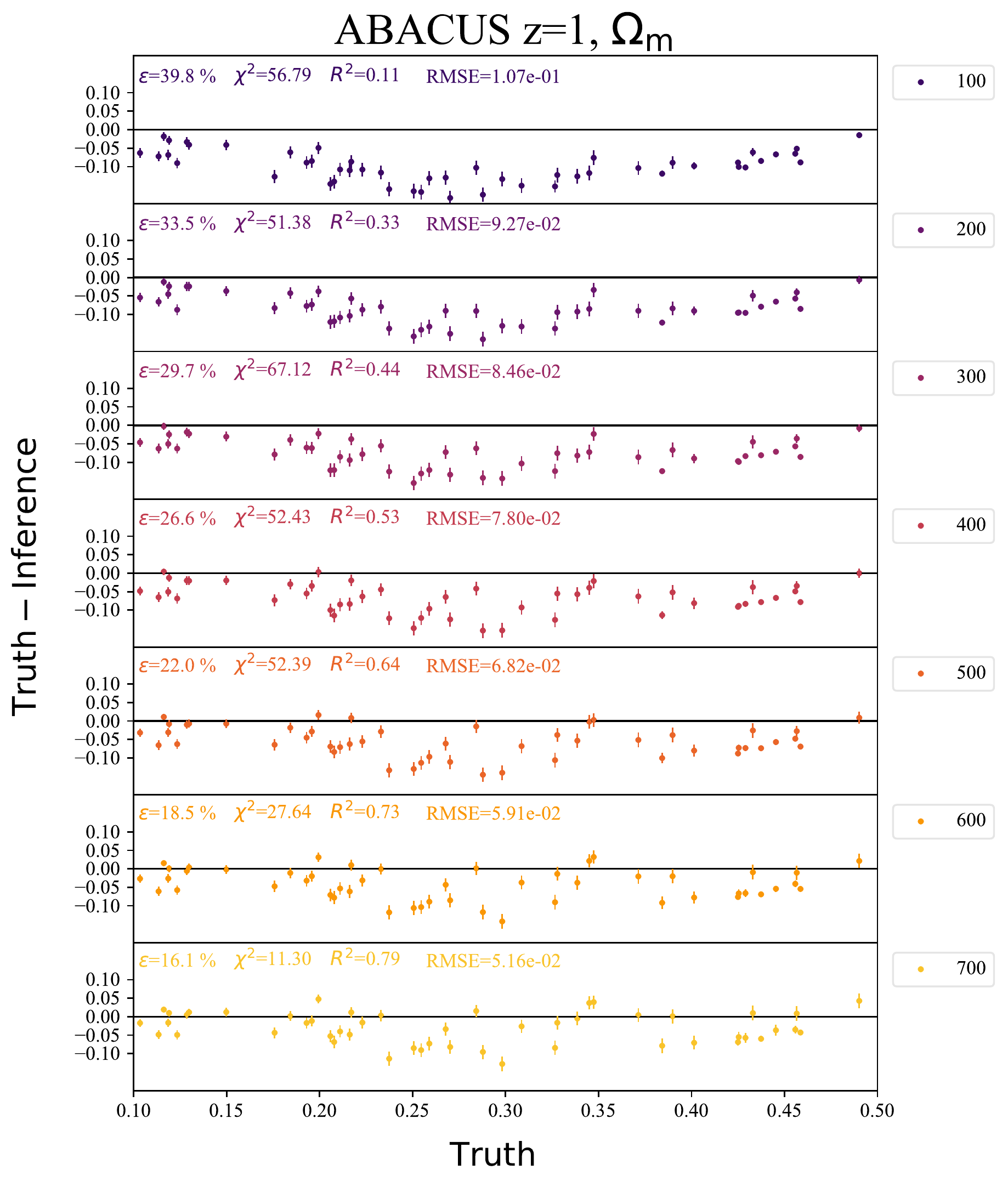}
    \includegraphics[width=.45\textwidth]{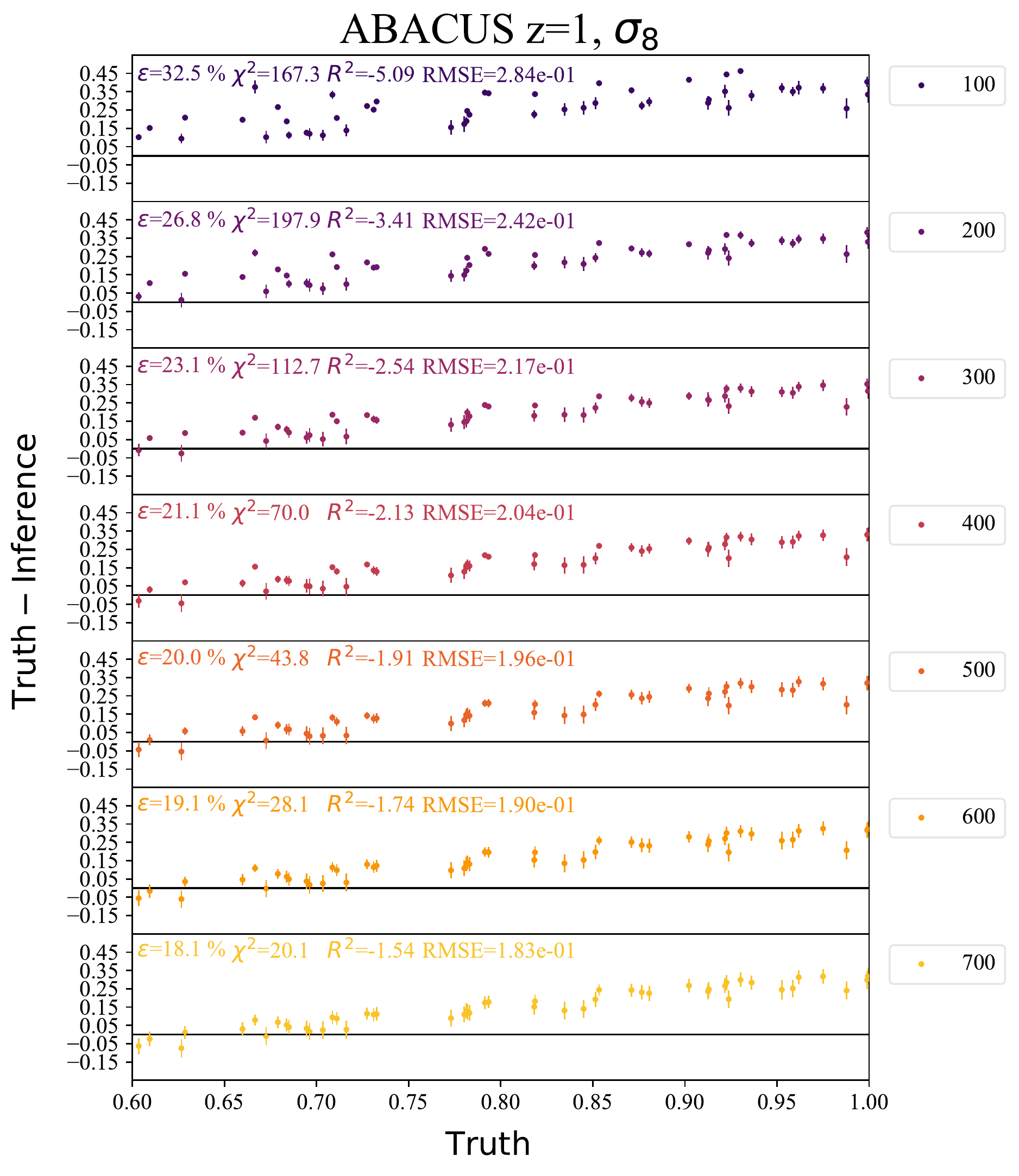}
    \caption{We evaluate the models trained to infer $\Omega_{\rm m}$ and $\sigma_{\rm 8}$ on Gadget simulations at redshift $z=1$ on halo catalogues from Abacus at redshift $z=1$. As it can be seen, the models are unable to predict $\Omega_{\rm m}$ and $\sigma_{\rm 8}$ for halos of all halo particle thresholds which is expected because the model did not receive any information regarding the redshift of the simulation and clustering changes with redshift. It is interesting to note that while $\Omega_{\rm m}$ is consistently over-predicted, the opposite is true for $\sigma_{\rm 8}$. However, for both parameters, the accuracies increase for larger particle thresholds.}
    \label{fig:z1}
\end{figure*}

We investigate whether the GNNs are able to extrapolate to redshifts higher than the one used in training, which was $z=0$. We find that the models are unable to predict the cosmological parameters at higher redshifts. For instance, when testing the GNN trained to predict $\Omega_{\rm m}$ on the Abacus catalogues at $z=1$, the predictions are significantly larger than the true values for all halo masses. This is shown in the left plot of Fig. \ref{fig:z1} where each panel is labeled with the halo particle threshold. This poor performance is expected since the models were not given information that distinguished the simulations based on redshift so they cannot learn how redshift affects the clustering of the halos. We also tested the model that was trained to infer $\sigma_{\rm 8}$ on Abacus simulations at $z=1$. Similar to $\Omega_{\rm m}$, the model fails for all particle thresholds. However, the results are under-predicted in this case. We do not attempt to explain these patterns in this work. Moreover, for both parameters, the predictions improve for catalogues generated with larger particle thresholds, which can possibly be explained by considering that the number density of halos for these catalogues are less severely affected by changes in redshift.

\section{Robustness to Halo Finder}\label{sec:subfind}
Here, we report the results of testing our models on halo catalogues generated by a halo finder that is different from the one used for training. Namely, we test the model on catalogues created with \textsc{SUBFIND} \citep{Subfind} on snapshots of the Gadget N-body simulations. 

\textsc{SUBFIND} works primarily by identifying local peaks in the three-dimensional density field and separating them by identifying a saddle point between them. Next, the overdensities and their surroundings are checked for gravitational self-boundness: those that are self-bound are registered as subhalos, and those that are not are attached to their neighboring overdensities, namely those they share saddle points with. \textsc{SUBFIND} operates on all particle types in the simulations, dark matter and baryonic alike.

To perform these tests, we consider the total mass of the halo contained in a sphere with a mean density that is 200 times the mean density of the Universe at redshift $z=0$. We find that our models are unable to accurately predict $\Omega_{\rm m}$ for catalogues created with any particle threshold, with relative errors of $\epsilon\sim15\%$. Similarly, for $\sigma_{\rm 8}$, the model predictions exhibit larger error for all particle thresholds. However, in this case, the relative error decreases with the particle threshold, from $\epsilon\sim 24\%$ to $\epsilon\sim 11\%$ as the particle threshold increases from 100 to 1000.

Upon further investigation, we find that the halo catalogues generated with \textsc{SUBFIND} exhibit significant differences in the halos spatial distribution and velocity fields compared to those generated using \textsc{Rockstar}. Specifically, when comparing the velocity modulus distribution curve for the 100 most massive halos from the two simulations, the velocity moduli of the \textsc{Rockstar} halos are noticeably smaller than those of the \textsc{SUBFIND} halos. Moreover, there are several halos whose positions differed between the two catalogues. These differences may affect the clustering of the halos and account for the inability of the model to accurately infer the cosmological parameters.

\section{Inferring \texorpdfstring{$\sigma_{\rm 8}$}{S8} for low mass halos}
\label{Sec:s8_mass_range}

\begin{figure*}[h!]
    \centering
    \includegraphics[width=1\textwidth]{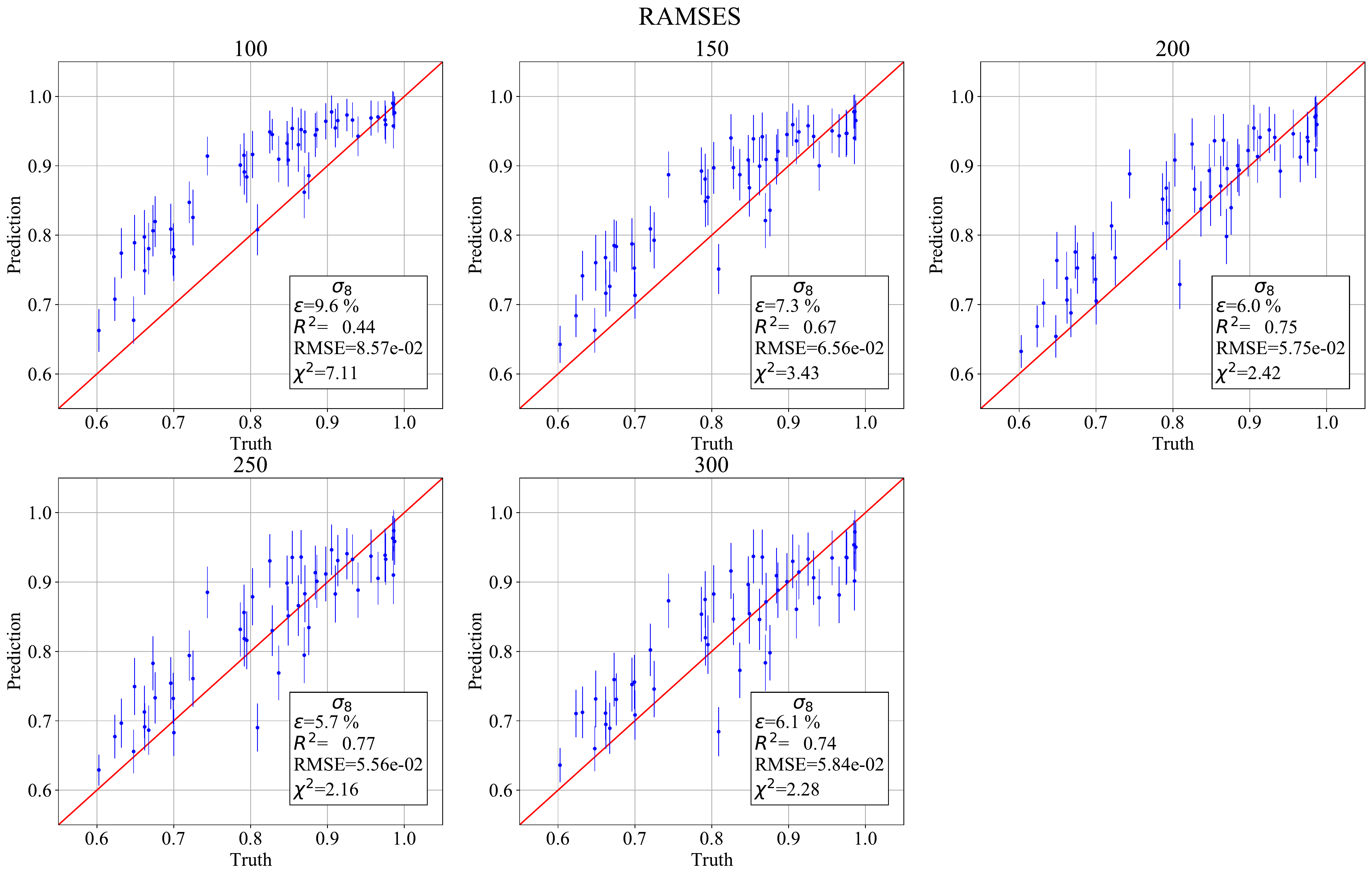}
    \caption{We train a GNN to infer $\sigma_{\rm 8}$ using halo catalogues of the Gadget N-body simulations created using random minimum particle thresholds ranging from [100, 300]. When we test this model on different N-body codes, such as Ramses, it is not robust for catalogues created using small particle thresholds. This is demonstrated by the first two plots of the figure, were the predictions are made for halos that contain at least 100 and 150 particles. An upwards offset from the truth can be seen for the predictions of these catalogues. Increasing the particle threshold improves the model performance and suggests that robustness can be found across the N-body simulations for larger halos.}
    \label{fig:RAMSES_sigma8_100_300}
\end{figure*}

Here we discuss the motivation behind choosing the range of minimum particle thresholds to be $[100,700]$ for $\sigma_{\rm 8}$ as outlined in Section \ref{subsec:inferring_sigma8}. When we train the GNN to infer $\sigma_{\rm 8}$ on halo catalogues that were created using particle thresholds ranging from $[100,500]$, we find that the model is able to achieve constraints with a mean relative error of $\epsilon = 4.6 \%$ and a Chi squared value of $\chi ^2 = 1.14$. However, when we evaluate this model on different N-body simulations, we find that the model performs poorly for halo catalogues constructed with small minimum particle thresholds. For instance, it can be seen in Fig. \ref{fig:RAMSES_sigma8_100_300}, when evaluated on Ramses, the model is only able to achieve $R^2 = 0.44$ and $R^2 = 0.67$ for halo catalogues created with minimum particle values of 100 and 150, respectively. The predictions also exhibit a significant offset from the truth. However, the model improves as the minimum particle threshold increases to 250 or 300, as shown in the figure. This trend suggests that field-level inference for $\sigma_{\rm 8}$ can be robust across different N-body simulations only for large halos. Hence, we decide to train the model on a larger interval of particle thresholds.

\section{Robustness on halo properties} \label{sec:both}

\begin{figure*}
    \centering
    \includegraphics[width=1\textwidth]{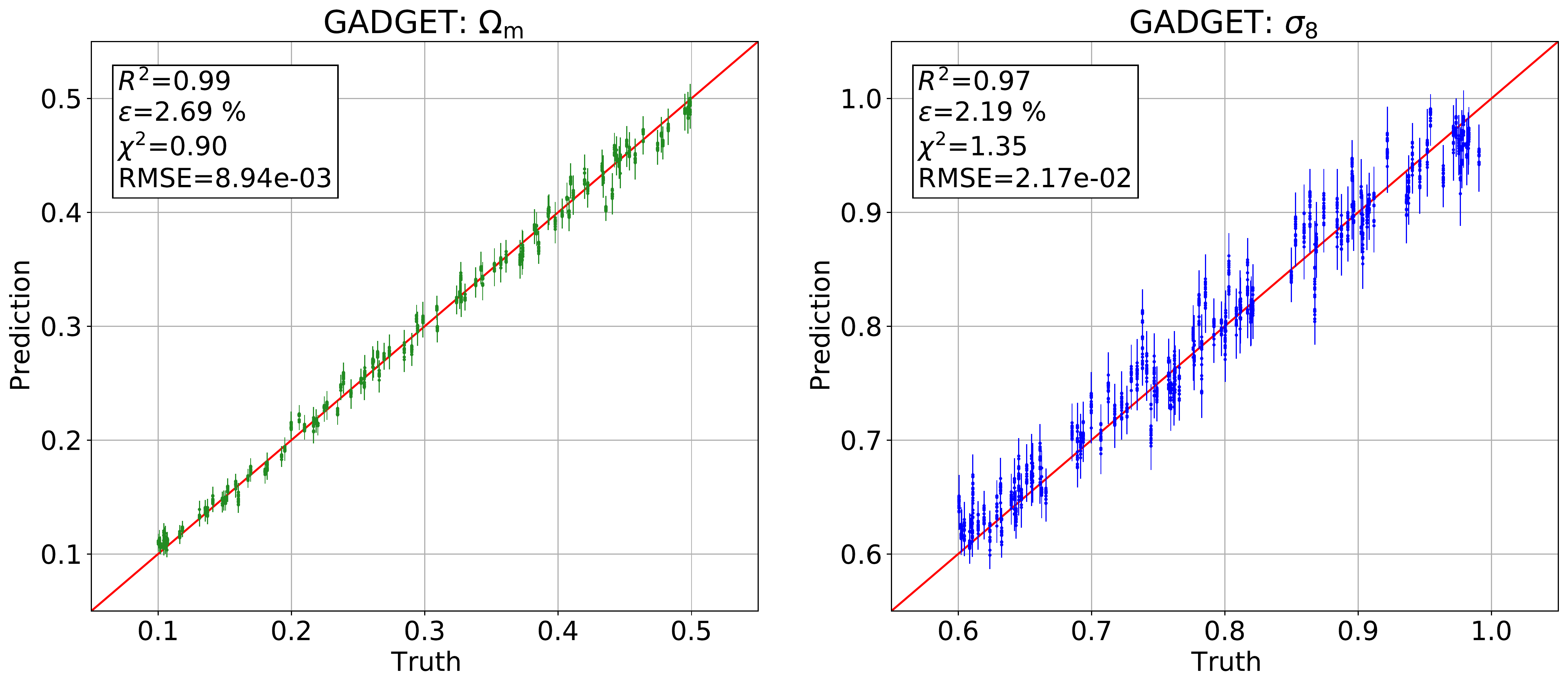}
    \caption{We train a GNN to infer $\Omega_{\rm m}$ and $\sigma_{\rm 8}$ simultaneously using halo catalogues from the Gadget N-body simulations. We employ the following halo properties as the initial node features: $V$, $c$, and $V_{\rm max}$. The plots illustrate the predictions of this model when evaluated on a test set of the Gadget simulations. As it can be seen, including the additional halo properties, $c$, and $V_{\rm max}$, significantly tightens the constraints on both $\Omega_{\rm m}$ and $\sigma_{\rm 8}$ compared to the previous accuracies of the models discussed in Section \ref{sec:results}. We emphasize that the test set on which this model is evaluated contains halo catalogues created with random halo particle thresholds ranging from [100, 1000], indicating that the model has marginalized over the number density of halos per catalogue.}
    \label{fig:gadget_both}
\end{figure*}

Here we describe in more detail the results of the network trained to infer both $\Omega_{\rm m}$ and $\sigma_8$ from catalogues that contain halo positions, velocity moduli, concentrations, and maximum circular velocities. We train this model on a larger interval of minimum halo particle thresholds, [100, 1000], to investigate the full extent of the model robustness.

\begin{figure*}
    \centering
    \includegraphics[width=1\textwidth]{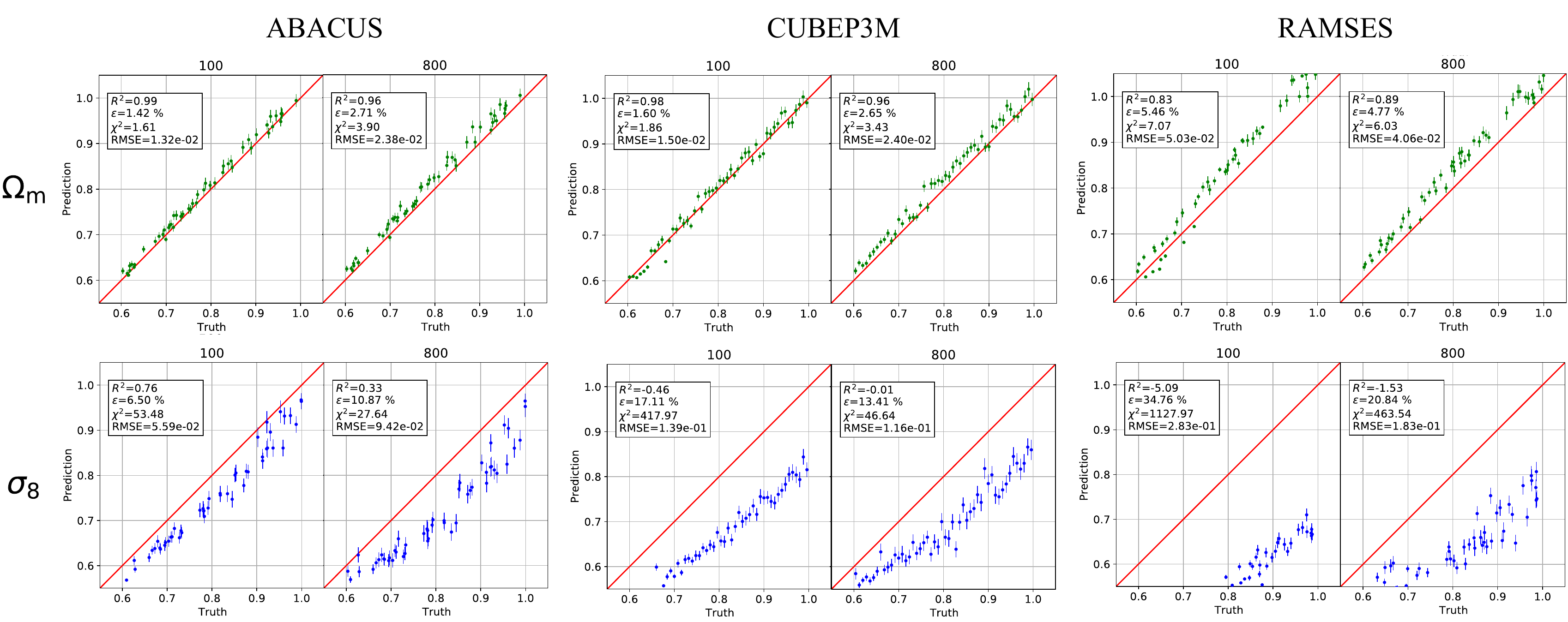}
    \caption{We test the model shown in Fig. \ref{fig:gadget_both} on the different N-body simulations: Abacus, CUBEP$^3$M, and Ramses. The first row of plots shows the model predictions for the parameter $\Omega_{\rm m}$ and the second row shows the predictions for $\sigma_{\rm 8}$. Clearly, the model is unable to extrapolate to the different simulations for either parameter. However, it is evident that the predictions for $\Omega_{\rm m}$ are closer to the truth than for $\sigma_{\rm 8}$.}
    \label{fig:simultaneous_nbody}
\end{figure*}

We show the evaluation of this model on a test set of the Gadget simulations in Fig. \ref{fig:gadget_both}. As can be seen, the inclusion of the additional halo properties ($c$ and $V_{\rm max}$) significantly increases the accuracy of the predictions for both $\Omega_{\rm m}$ and $\sigma_{\rm 8}$ compared to the accuracies of the models used to infer $\Omega_{\rm m}$ and $\sigma_8$ individually. This model is able to predict $\Omega_{\rm m}$ with a relative error of $\epsilon = 2.7 \%$ and a coefficient of determination of $R^2 = 0.99$. For $\sigma_{\rm 8}$, the relative error is $\epsilon = 2.19 \%$ and a coefficient of determination of $R^2 = 0.97$. Moreover, both predictions exhibit Chi squared values that are close to one, $\chi^2 = 0.90$ and $\chi^2 = 1.32$, which indicate that the standard deviations of the posteriors are being accurately inferred as well. 

However, this model is not robust across different N-body simulations in predicting both $\Omega_{\rm m}$ and $\sigma_{\rm 8}$. When we test this model on the different N-body simulations, all the predictions for $\Omega_{\rm m}$ exhibit an overall upward offset from the truth for all particle thresholds. Similarly, the predictions for $\sigma_{\rm 8}$ consistently display a large downward offset from the truth for all simulations and particle thresholds. In Fig. \ref{fig:simultaneous_nbody}, we show these results for a subset of the simulations, namely Abacus, CUBEP$^3$M, and Ramses. We also only include the plots for the particle thresholds of 100 and 800 to demonstrate the lack of accuracy regardless of the minimum halo particle thresholds, however this pattern is consistently seen across all other particle thresholds. A possible explanation for this lack of robustness is that the input properties, $c$ and $V_{\rm max}$, are more dependent on the interior of halos which can more easily vary among the different simulation codes as opposed to global halo properties such as their masses and positions. However, it is evident that the predictions for $\Omega_{\rm m}$ are closer to the truth than for $\sigma_{\rm 8}$, as it can be seen in the plots. While this suggests that $\Omega_{\rm m}$ is a more stable parameter across the different N-body codes, we do not attempt to explain these results here.

\bibliography{references}{}

\begin{thebibliography}{}
\expandafter\ifx\csname natexlab\endcsname\relax\def\natexlab#1{#1}\fi
\providecommand{\url}[1]{\href{#1}{#1}}
\providecommand{\dodoi}[1]{doi:~\href{http://doi.org/#1}{\nolinkurl{#1}}}
\providecommand{\doeprint}[1]{\href{http://ascl.net/#1}{\nolinkurl{http://ascl.net/#1}}}
\providecommand{\doarXiv}[1]{\href{https://arxiv.org/abs/#1}{\nolinkurl{https://arxiv.org/abs/#1}}}

\bibitem[{Akiba {et~al.}(2019)Akiba, Sano, Yanase, Ohta, \& Koyama}]{Optuna}
Akiba, T., Sano, S., Yanase, T., Ohta, T., \& Koyama, M. 2019, in Proceedings
  of the 25rd {ACM} {SIGKDD} International Conference on Knowledge Discovery
  and Data Mining

\bibitem[{Battaglia {et~al.}(2018)Battaglia, Hamrick, Bapst,
  Sanchez{-}Gonzalez, Zambaldi, Malinowski, Tacchetti, Raposo, Santoro,
  Faulkner, G{\"{u}}l{\c{c}}ehre, Song, Ballard, Gilmer, Dahl, Vaswani, Allen,
  Nash, Langston, Dyer, Heess, Wierstra, Kohli, Botvinick, Vinyals, Li, \&
  Pascanu}]{battaglia}
Battaglia, P.~W., Hamrick, J.~B., Bapst, V., {et~al.} 2018, CoRR,
  abs/1806.01261

\bibitem[{{Beck} {et~al.}(2016){Beck}, {Murante}, {Arth}, {Remus}, {Teklu},
  {Donnert}, {Planelles}, {Beck}, {F{\"o}rster}, {Imgrund}, {Dolag}, \&
  {Borgani}}]{Beck2016}
{Beck}, A.~M., {Murante}, G., {Arth}, A., {et~al.} 2016, \mnras, 455, 2110,
  \dodoi{10.1093/mnras/stv2443}

\bibitem[{{Behroozi} {et~al.}(2013){Behroozi}, {Wechsler}, \& {Wu}}]{rockstar}
{Behroozi}, P.~S., {Wechsler}, R.~H., \& {Wu}, H.-Y. 2013, \apj, 762, 109,
  \dodoi{10.1088/0004-637X/762/2/109}

\bibitem[{{Bird} {et~al.}(2022){Bird}, {Ni}, {Di Matteo}, {Croft}, {Feng}, \&
  {Chen}}]{Astrid2}
{Bird}, S., {Ni}, Y., {Di Matteo}, T., {et~al.} 2022, \mnras, 512, 3703,
  \dodoi{10.1093/mnras/stac648}

\bibitem[{Bronstein {et~al.}(2021)Bronstein, Bruna, Cohen, \&
  Velickovic}]{bronstein}
Bronstein, M.~M., Bruna, J., Cohen, T., \& Velickovic, P. 2021, CoRR,
  abs/2104.13478

\bibitem[{{Bryan} {et~al.}(2014){Bryan}, {Norman}, {O'Shea}, {Abel}, {Wise},
  {Turk}, {Reynolds}, {Collins}, {Wang}, {Skillman}, {Smith}, {Harkness},
  {Bordner}, {Kim}, {Kuhlen}, {Xu}, {Goldbaum}, {Hummels}, {Kritsuk}, {Tasker},
  {Skory}, {Simpson}, {Hahn}, {Oishi}, {So}, {Zhao}, {Cen}, {Li}, \& {Enzo
  Collaboration}}]{Enzo}
{Bryan}, G.~L., {Norman}, M.~L., {O'Shea}, B.~W., {et~al.} 2014, \apjs, 211,
  19, \dodoi{10.1088/0067-0049/211/2/19}

\bibitem[{{Dav{\'e}} {et~al.}(2019){Dav{\'e}}, {Angl{\'e}s-Alc{\'a}zar},
  {Narayanan}, {Li}, {Rafieferantsoa}, \& {Appleby}}]{Dave2019_Simba}
{Dav{\'e}}, R., {Angl{\'e}s-Alc{\'a}zar}, D., {Narayanan}, D., {et~al.} 2019,
  \mnras, 486, 2827, \dodoi{10.1093/mnras/stz937}

\bibitem[{{DESI Collaboration} {et~al.}(2016){DESI Collaboration}, {Aghamousa},
  {Aguilar}, {Ahlen}, {Alam}, {et~al.}}]{DESI_2016}
{DESI Collaboration}, {Aghamousa}, A., {Aguilar}, J., {et~al.} 2016, arXiv
  e-prints, arXiv:1611.00036.
\newblock \doarXiv{1611.00036}

\bibitem[{{Dolag} {et~al.}(2004){Dolag}, {Jubelgas}, {Springel}, {Borgani}, \&
  {Rasia}}]{Dolag2004}
{Dolag}, K., {Jubelgas}, M., {Springel}, V., {Borgani}, S., \& {Rasia}, E.
  2004, \apjl, 606, L97, \dodoi{10.1086/420966}

\bibitem[{{Euclid Collaboration} {et~al.}(2022){Euclid Collaboration},
  {Castro}, {Fumagalli}, {Angulo}, {Bocquet}, {Borgani}, {Carbone}, {Dakin},
  {Dolag}, {Giocoli}, {Monaco}, {Ragagnin}, {Saro}, {Sefusatti}, {Costanzi},
  {Amara}, {Amendola}, {Baldi}, {Bender}, {Bodendorf}, {Branchini}, {Brescia},
  {Camera}, {Capobianco}, {Carretero}, {Castellano}, {Cavuoti}, {Cimatti},
  {Cledassou}, {Congedo}, {Conversi}, {Copin}, {Corcione}, {Courbin}, {Da
  Silva}, {Degaudenzi}, {Douspis}, {Dubath}, {Duncan}, {Dupac}, {Farrens},
  {Ferriol}, {Fosalba}, {Frailis}, {Franceschi}, {Galeotta}, {Garilli},
  {Gillis}, {Grazian}, {Gruppi}, {Haugan}, {Hormuth}, {Hornstrup}, {Hudelot},
  {Jahnke}, {Kermiche}, {Kitching}, {Kunz}, {Kurki-Suonio}, {Lilje}, {Lloro},
  {Mansutti}, {Marggraf}, {Meneghetti}, {Merlin}, {Meylan}, {Moresco},
  {Moscardini}, {Munari}, {Niemi}, {Padilla}, {Paltani}, {Pasian}, {Pedersen},
  {Pettorino}, {Pires}, {Polenta}, {Poncet}, {Popa}, {Pozzetti}, {Raison},
  {Rebolo}, {Renzi}, {Rhodes}, {Riccio}, {Romelli}, {Saglia}, {Sapone},
  {Sartoris}, {Schneider}, {Seidel}, {Sirri}, {Stanco}, {Tallada Cresp{\'\i}},
  {Taylor}, {Toledo-Moreo}, {Torradeflot}, {Tutusaus}, {Valentijn},
  {Valenziano}, {Vassallo}, {Wang}, {Weller}, {Zacchei}, {Zamorani}, {Andreon},
  {Bardelli}, {Bozzo}, {Colodro-Conde}, {Di Ferdinando}, {Farina},
  {Graci{\'a}-Carpio}, {Lindholm}, {Neissner}, {Scottez}, {Tenti}, {Zucca},
  {Baccigalupi}, {Balaguera-Antol{\'\i}nez}, {Ballardini}, {Bernardeau},
  {Biviano}, {Blanchard}, {Borlaff}, {Burigana}, {Cabanac}, {Cappi},
  {Carvalho}, {Casas}, {Castignani}, {Cooray}, {Coupon}, {Courtois}, {Davini},
  {De Lucia}, {Desprez}, {Dole}, {Escartin}, {Escoffier}, {Finelli}, {Ganga},
  {Garcia-Bellido}, {George}, {Gozaliasl}, {Hildebrandt}, {Hook}, {Ili{\'c}},
  {Kansal}, {Keihanen}, {Kirkpatrick}, {Loureiro}, {Macias-Perez},
  {Magliocchetti}, {Maoli}, {Marcin}, {Martinelli}, {Martinet}, {Matthew},
  {Maturi}, {Metcalf}, {Morgante}, {Nadathur}, {Nucita}, {Patrizii}, {Peel},
  {Popa}, {Porciani}, {Potter}, {Pourtsidou}, {P{\"o}ntinen}, {S{\'a}nchez},
  {Sakr}, {Schirmer}, {Sereno}, {Spurio Mancini}, {Teyssier}, {Valiviita},
  {Veropalumbo}, \& {Viel}}]{calibration_of_hmf}
{Euclid Collaboration}, {Castro}, T., {Fumagalli}, A., {et~al.} 2022, arXiv
  e-prints, arXiv:2208.02174.
\newblock \doarXiv{2208.02174}

\bibitem[{Fey \& Lenssen(2019)}]{Fey_Fast_Graph_Representation_2019}
Fey, M., \& Lenssen, J.~E. 2019, {Fast Graph Representation Learning with
  PyTorch Geometric}, 2.0.2.
\newblock \url{https://github.com/pyg-team/pytorch_geometric}

\bibitem[{{Fluri} {et~al.}(2019){Fluri}, {Kacprzak}, {Lucchi}, {Refregier},
  {Amara}, {Hofmann}, \& {Schneider}}]{Fluri_19}
{Fluri}, J., {Kacprzak}, T., {Lucchi}, A., {et~al.} 2019, \prd, 100, 063514,
  \dodoi{10.1103/PhysRevD.100.063514}

\bibitem[{{Garrison} {et~al.}(2021{\natexlab{a}}){Garrison}, {Eisenstein},
  {Ferrer}, {Maksimova}, \& {Pinto}}]{2021MNRAS.508..575G}
{Garrison}, L.~H., {Eisenstein}, D.~J., {Ferrer}, D., {Maksimova}, N.~A., \&
  {Pinto}, P.~A. 2021{\natexlab{a}}, \mnras, 508, 575,
  \dodoi{10.1093/mnras/stab2482}

\bibitem[{{Garrison} {et~al.}(2019){Garrison}, {Eisenstein}, \&
  {Pinto}}]{2019MNRAS.485.3370G}
{Garrison}, L.~H., {Eisenstein}, D.~J., \& {Pinto}, P.~A. 2019, \mnras, 485,
  3370, \dodoi{10.1093/mnras/stz634}

\bibitem[{{Garrison} {et~al.}(2021{\natexlab{b}}){Garrison}, {Joyce}, \&
  {Eisenstein}}]{2021MNRAS.504.3550G}
{Garrison}, L.~H., {Joyce}, M., \& {Eisenstein}, D.~J. 2021{\natexlab{b}},
  \mnras, 504, 3550, \dodoi{10.1093/mnras/stab1096}

\bibitem[{{Greengard} \& {Rokhlin}(1987)}]{Greengard1987}
{Greengard}, L., \& {Rokhlin}, V. 1987, Journal of Computational Physics, 73,
  325, \dodoi{10.1016/0021-9991(87)90140-9}

\bibitem[{Gupta {et~al.}(2018)Gupta, Matilla, Hsu, \& Haiman}]{Gupta_18}
Gupta, A., Matilla, J. M.~Z., Hsu, D., \& Haiman, Z. 2018, Phys. Rev. D, 97,
  103515, \dodoi{10.1103/PhysRevD.97.103515}

\bibitem[{Hamilton(2020)}]{hamilton}
Hamilton, W.~L. 2020, Synthesis Lectures on Artificial Intelligence and Machine
  Learning, 14, 1

\bibitem[{{Harnois-D{\'e}raps} {et~al.}(2013){Harnois-D{\'e}raps}, {Pen},
  {Iliev}, {Merz}, {Emberson}, \& {Desjacques}}]{2013MNRAS.436..540H}
{Harnois-D{\'e}raps}, J., {Pen}, U.-L., {Iliev}, I.~T., {et~al.} 2013, \mnras,
  436, 540, \dodoi{10.1093/mnras/stt1591}

\bibitem[{Heitmann {et~al.}(2005)Heitmann, Ricker, Warren, \&
  Habib}]{Heitmann_2005}
Heitmann, K., Ricker, P.~M., Warren, M.~S., \& Habib, S. 2005, The
  Astrophysical Journal Supplement Series, 160, 28, \dodoi{10.1086/432646}

\bibitem[{Heitmann {et~al.}(2008)Heitmann, Luki{\'{c}}, Fasel, Habib, Warren,
  White, Ahrens, Ankeny, Armstrong, O{\textquotesingle}Shea, Ricker, Springel,
  Stadel, \& Trac}]{Heitmann_2008}
Heitmann, K., Luki{\'{c}}, Z., Fasel, P., {et~al.} 2008, Computational Science
  {\&} Discovery, 1, 015003, \dodoi{10.1088/1749-4699/1/1/015003}

\bibitem[{{Hirschmann} {et~al.}(2014){Hirschmann}, {Dolag}, {Saro}, {Bachmann},
  {Borgani}, \& {Burkert}}]{Hirschmann2014}
{Hirschmann}, M., {Dolag}, K., {Saro}, A., {et~al.} 2014, \mnras, 442, 2304,
  \dodoi{10.1093/mnras/stu1023}

\bibitem[{{Hockney} \& {Eastwood}(1988)}]{Hockney_1998}
{Hockney}, R.~W., \& {Eastwood}, J.~W. 1988, {Computer simulation using
  particles}

\bibitem[{{Hopkins}(2015)}]{Gizmo}
{Hopkins}, P.~F. 2015, \mnras, 450, 53, \dodoi{10.1093/mnras/stv195}

\bibitem[{Inman \& Ali-Ha\"{\i}moud(2019)}]{PhysRevD.100.083528}
Inman, D., \& Ali-Ha\"{\i}moud, Y. 2019, Phys. Rev. D, 100, 083528,
  \dodoi{10.1103/PhysRevD.100.083528}

\bibitem[{{Jeffrey} {et~al.}(2020){Jeffrey}, {Alsing}, \&
  {Lanusse}}]{Niall_2020}
{Jeffrey}, N., {Alsing}, J., \& {Lanusse}, F. 2020, arXiv e-prints,
  arXiv:2009.08459.
\newblock \doarXiv{2009.08459}

\bibitem[{Jeffrey \& Wandelt(2020)}]{jeffrey_wandelt}
Jeffrey, N., \& Wandelt, B.~D. 2020, in {34th Conference on Neural Information
  Processing Systems}, Online Conference, Canada.
\newblock \url{https://hal.archives-ouvertes.fr/hal-03047530}

\bibitem[{{Jubelgas} {et~al.}(2004){Jubelgas}, {Springel}, \&
  {Dolag}}]{Jubelgas2004}
{Jubelgas}, M., {Springel}, V., \& {Dolag}, K. 2004, \mnras, 351, 423,
  \dodoi{10.1111/j.1365-2966.2004.07801.x}

\bibitem[{{Laureijs} {et~al.}(2011){Laureijs}, {Amiaux}, {Arduini},
  {Augu{\`e}res}, {Brinchmann}, {et~al.}}]{Laureijs_2011}
{Laureijs}, R., {Amiaux}, J., {Arduini}, S., {et~al.} 2011, arXiv e-prints,
  arXiv:1110.3193.
\newblock \doarXiv{1110.3193}

\bibitem[{{Loshchilov} \& {Hutter}(2017)}]{AdamW}
{Loshchilov}, I., \& {Hutter}, F. 2017, arXiv e-prints, arXiv:1711.05101.
\newblock \doarXiv{1711.05101}

\bibitem[{{LSST Science Collaboration} {et~al.}(2009){LSST Science
  Collaboration}, {Abell}, {Allison}, {Anderson}, {Andrew},
  {et~al.}}]{LSST_2009}
{LSST Science Collaboration}, {Abell}, P.~A., {Allison}, J., {et~al.} 2009,
  arXiv e-prints, arXiv:0912.0201.
\newblock \doarXiv{0912.0201}

\bibitem[{{Makinen} {et~al.}(2022){Makinen}, {Charnock}, {Lemos}, {Porqueres},
  {Heavens}, \& {Wandelt}}]{Makinen_2022}
{Makinen}, T.~L., {Charnock}, T., {Lemos}, P., {et~al.} 2022, arXiv e-prints,
  arXiv:2207.05202.
\newblock \doarXiv{2207.05202}

\bibitem[{{Metchnik}(2009)}]{2009PhDT.......175M}
{Metchnik}, M. V.~L. 2009, PhD thesis, University of Arizona

\bibitem[{{Ni} {et~al.}(2022){Ni}, {Di Matteo}, {Bird}, {Croft}, {Feng},
  {Chen}, {Tremmel}, {DeGraf}, \& {Li}}]{Astrid1}
{Ni}, Y., {Di Matteo}, T., {Bird}, S., {et~al.} 2022, \mnras, 513, 670,
  \dodoi{10.1093/mnras/stac351}

\bibitem[{{Ntampaka} {et~al.}(2019){Ntampaka}, {Eisenstein}, {Yuan}, \&
  {Garrison}}]{Ntampaka_19}
{Ntampaka}, M., {Eisenstein}, D.~J., {Yuan}, S., \& {Garrison}, L.~H. 2019,
  arXiv e-prints, arXiv:1909.10527.
\newblock \doarXiv{1909.10527}

\bibitem[{{Paszke} {et~al.}(2019){Paszke}, {Gross}, {Massa}, {Lerer},
  {Bradbury}, {Chanan}, {Killeen}, {Lin}, {Gimelshein}, {Antiga}, {Desmaison},
  {K{\"o}pf}, {Yang}, {DeVito}, {Raison}, {Tejani}, {Chilamkurthy}, {Steiner},
  {Fang}, {Bai}, \& {Chintala}}]{Pytorch}
{Paszke}, A., {Gross}, S., {Massa}, F., {et~al.} 2019, arXiv e-prints,
  arXiv:1912.01703.
\newblock \doarXiv{1912.01703}

\bibitem[{{Pillepich} {et~al.}(2018){Pillepich}, {Springel}, {Nelson}, {Genel},
  {Naiman}, {Pakmor}, {Hernquist}, {Torrey}, {Vogelsberger}, {Weinberger}, \&
  {Marinacci}}]{PillepichA_16a}
{Pillepich}, A., {Springel}, V., {Nelson}, D., {et~al.} 2018, \mnras, 473,
  4077, \dodoi{10.1093/mnras/stx2656}

\bibitem[{{Potter} {et~al.}(2017){Potter}, {Stadel}, \& {Teyssier}}]{PKDGrav}
{Potter}, D., {Stadel}, J., \& {Teyssier}, R. 2017, Computational Astrophysics
  and Cosmology, 4, 2, \dodoi{10.1186/s40668-017-0021-1}

\bibitem[{{Predehl} {et~al.}(2021){Predehl}, {Andritschke}, {Arefiev},
  {Babyshkin}, {Batanov}, {et~al.}}]{Predhel_2021}
{Predehl}, P., {Andritschke}, R., {Arefiev}, V., {et~al.} 2021, \aap, 647, A1,
  \dodoi{10.1051/0004-6361/202039313}

\bibitem[{{Ribli} {et~al.}(2019){Ribli}, {Pataki}, {Zorrilla Matilla}, {Hsu},
  {Haiman}, \& {Csabai}}]{Ribli_19}
{Ribli}, D., {Pataki}, B.~{\'A}., {Zorrilla Matilla}, J.~M., {et~al.} 2019,
  \mnras, 490, 1843, \dodoi{10.1093/mnras/stz2610}

\bibitem[{{Schmelzle} {et~al.}(2017){Schmelzle}, {Lucchi}, {Kacprzak}, {Amara},
  {Sgier}, {R{\'e}fr{\'e}gier}, \& {Hofmann}}]{Schmelzle_17}
{Schmelzle}, J., {Lucchi}, A., {Kacprzak}, T., {et~al.} 2017, arXiv e-prints,
  arXiv:1707.05167.
\newblock \doarXiv{1707.05167}

\bibitem[{Schneider {et~al.}(2016)Schneider, Teyssier, Potter, Stadel, Reed,
  Onions, Pearce, Smith, Springel, \& Scoccimarro}]{schneider}
Schneider, A., Teyssier, R., Potter, D., {et~al.} 2016, Journal of Cosmology
  and Astroparticle Physics, 2016, \dodoi{10.1088/1475-7516/2016/04/047}

\bibitem[{{Spergel} {et~al.}(2015){Spergel}, {Gehrels}, {Baltay}, {Bennett},
  {Breckinridge}, {Donahue}, {et~al.}}]{Spergel_2015}
{Spergel}, D., {Gehrels}, N., {Baltay}, C., {et~al.} 2015, arXiv e-prints,
  arXiv:1503.03757.
\newblock \doarXiv{1503.03757}

\bibitem[{Springel(2005)}]{Gadget}
Springel, V. 2005, Mon. Not. Roy. Astron. Soc., 364, 1105,
  \dodoi{10.1111/j.1365-2966.2005.09655.x}

\bibitem[{{Springel}(2010)}]{Arepo}
{Springel}, V. 2010, \mnras, 401, 791, \dodoi{10.1111/j.1365-2966.2009.15715.x}

\bibitem[{{Springel} {et~al.}(2001){Springel}, {White}, {Tormen}, \&
  {Kauffmann}}]{Subfind}
{Springel}, V., {White}, S.~D.~M., {Tormen}, G., \& {Kauffmann}, G. 2001,
  \mnras, 328, 726, \dodoi{10.1046/j.1365-8711.2001.04912.x}

\bibitem[{{Takada} {et~al.}(2014){Takada}, {Ellis}, {Chiba}, {Greene},
  {Aihara}, {et~al.}}]{Takada_2014}
{Takada}, M., {Ellis}, R.~S., {Chiba}, M., {et~al.} 2014, \pasj, 66, R1,
  \dodoi{10.1093/pasj/pst019}

\bibitem[{{Taylor} \& {Braun}(1999)}]{Taylor_1999}
{Taylor}, A.~R., \& {Braun}, R. 1999, in Science with the Square Kilometer
  Array : a Next Generation World Radio Observatory

\bibitem[{{Teyssier}(2002)}]{Ramses}
{Teyssier}, R. 2002, \aap, 385, 337, \dodoi{10.1051/0004-6361:20011817}

\bibitem[{{Villaescusa-Navarro} {et~al.}(2020){Villaescusa-Navarro}, {Hahn},
  {Massara}, {Banerjee}, {Delgado}, {Ramanah}, {Charnock}, {Giusarma}, {Li},
  {Allys}, {Brochard}, {Uhlemann}, {Chiang}, {He}, {Pisani}, {Obuljen}, {Feng},
  {Castorina}, {Contardo}, {Kreisch}, {Nicola}, {Alsing}, {Scoccimarro},
  {Verde}, {Viel}, {Ho}, {Mallat}, {Wandelt}, \& {Spergel}}]{Quijote}
{Villaescusa-Navarro}, F., {Hahn}, C., {Massara}, E., {et~al.} 2020, \apjs,
  250, 2, \dodoi{10.3847/1538-4365/ab9d82}

\bibitem[{{Villaescusa-Navarro}
  {et~al.}(2021{\natexlab{a}}){Villaescusa-Navarro}, {Angl{\'e}s-Alc{\'a}zar},
  {Genel}, {Spergel}, {Li}, {Wandelt}, {Nicola}, {Thiele}, {Hassan}, {Zorrilla
  Matilla}, {Narayanan}, {Dave}, \& {Vogelsberger}}]{Paco_2021a}
{Villaescusa-Navarro}, F., {Angl{\'e}s-Alc{\'a}zar}, D., {Genel}, S., {et~al.}
  2021{\natexlab{a}}, arXiv e-prints, arXiv:2109.09747.
\newblock \doarXiv{2109.09747}

\bibitem[{{Villaescusa-Navarro}
  {et~al.}(2021{\natexlab{b}}){Villaescusa-Navarro}, {Genel}, {Angles-Alcazar},
  {Spergel}, {Li}, {Wandelt}, {Thiele}, {Nicola}, {Zorrilla Matilla}, {Shao},
  {Hassan}, {Narayanan}, {Dave}, \& {Vogelsberger}}]{Paco_2021b}
{Villaescusa-Navarro}, F., {Genel}, S., {Angles-Alcazar}, D., {et~al.}
  2021{\natexlab{b}}, arXiv e-prints, arXiv:2109.10360.
\newblock \doarXiv{2109.10360}

\bibitem[{{Villaescusa-Navarro}
  {et~al.}(2021{\natexlab{c}}){Villaescusa-Navarro}, {Angl{\'e}s-Alc{\'a}zar},
  {Genel}, {Spergel}, {Somerville}, {Dave}, {Pillepich}, {Hernquist}, {Nelson},
  {Torrey}, {Narayanan}, {Li}, {Philcox}, {La Torre}, {Maria Delgado}, {Ho},
  {Hassan}, {Burkhart}, {Wadekar}, {Battaglia}, {Contardo}, \&
  {Bryan}}]{villaescusanavarro2020camels}
{Villaescusa-Navarro}, F., {Angl{\'e}s-Alc{\'a}zar}, D., {Genel}, S., {et~al.}
  2021{\natexlab{c}}, \apj, 915, 71, \dodoi{10.3847/1538-4357/abf7ba}

\bibitem[{{Villaescusa-Navarro}
  {et~al.}(2022{\natexlab{a}}){Villaescusa-Navarro}, {Genel},
  {Angl{\'e}s-Alc{\'a}zar}, {Perez}, {Villanueva-Domingo}, {Wadekar}, {Shao},
  {Mohammad}, {Hassan}, {Moser}, {Lau}, {Machado Poletti Valle}, {Nicola},
  {Thiele}, {Jo}, {Philcox}, {Oppenheimer}, {Tillman}, {Hahn}, {Kaushal},
  {Pisani}, {Gebhardt}, {Delgado}, {Caliendo}, {Kreisch}, {Wong}, {Coulton},
  {Eickenberg}, {Parimbelli}, {Ni}, {Steinwandel}, {La Torre}, {Dave},
  {Battaglia}, {Nagai}, {Spergel}, {Hernquist}, {Burkhart}, {Narayanan},
  {Wandelt}, {Somerville}, {Bryan}, {Viel}, {Li}, {Irsic}, {Kraljic}, \&
  {Vogelsberger}}]{CAMELS_public}
{Villaescusa-Navarro}, F., {Genel}, S., {Angl{\'e}s-Alc{\'a}zar}, D., {et~al.}
  2022{\natexlab{a}}, arXiv e-prints, arXiv:2201.01300.
\newblock \doarXiv{2201.01300}

\bibitem[{{Villaescusa-Navarro}
  {et~al.}(2022{\natexlab{b}}){Villaescusa-Navarro}, {Genel},
  {Angl{\'e}s-Alc{\'a}zar}, {Thiele}, {Dave}, {Narayanan}, {Nicola}, {Li},
  {Villanueva-Domingo}, {Wandelt}, {Spergel}, {Somerville}, {Zorrilla Matilla},
  {Mohammad}, {Hassan}, {Shao}, {Wadekar}, {Eickenberg}, {Wong}, {Contardo},
  {Jo}, {Moser}, {Lau}, {Machado Poletti Valle}, {Perez}, {Nagai}, {Battaglia},
  \& {Vogelsberger}}]{CMD}
---. 2022{\natexlab{b}}, \apjs, 259, 61, \dodoi{10.3847/1538-4365/ac5ab0}

\bibitem[{Villanueva-Domingo(2022)}]{pablo_villanueva_domingo_2022_6485804}
Villanueva-Domingo, P. 2022, PabloVD/CosmoGraphNet, v1.0,  Zenodo,
  \dodoi{10.5281/zenodo.6485804}

\bibitem[{{Villanueva-Domingo} \& {Villaescusa-Navarro}(2022)}]{pablo}
{Villanueva-Domingo}, P., \& {Villaescusa-Navarro}, F. 2022, arXiv e-prints,
  arXiv:2204.13713.
\newblock \doarXiv{2204.13713}

\bibitem[{Weinberger {et~al.}(2020)Weinberger, Springel, \&
  Pakmor}]{Arepo_public}
Weinberger, R., Springel, V., \& Pakmor, R. 2020, Astrophys. J. Suppl., 248,
  32, \dodoi{10.3847/1538-4365/ab908c}

\bibitem[{{Weinberger} {et~al.}(2017){Weinberger}, {Springel}, {Hernquist},
  {Pillepich}, {Marinacci}, {Pakmor}, {Nelson}, {Genel}, {Vogelsberger},
  {Naiman}, \& {Torrey}}]{WeinbergerR_16a}
{Weinberger}, R., {Springel}, V., {Hernquist}, L., {et~al.} 2017, \mnras, 465,
  3291, \dodoi{10.1093/mnras/stw2944}

\bibitem[{Zaheer {et~al.}(2017)Zaheer, Kottur, Ravanbakhsh, Poczos,
  Salakhutdinov, \& Smola}]{deepset}
Zaheer, M., Kottur, S., Ravanbakhsh, S., {et~al.} 2017, in Advances in Neural
  Information Processing Systems, ed. I.~Guyon, U.~V. Luxburg, S.~Bengio,
  H.~Wallach, R.~Fergus, S.~Vishwanathan, \& R.~Garnett, Vol.~30 (Curran
  Associates, Inc.).
\newblock
  \url{https://proceedings.neurips.cc/paper/2017/file/f22e4747da1aa27e363d86d40ff442fe-Paper.pdf}

\bibitem[{{Zorrilla Matilla} {et~al.}(2020){Zorrilla Matilla}, {Sharma}, {Hsu},
  \& {Haiman}}]{Jose_2020}
{Zorrilla Matilla}, J.~M., {Sharma}, M., {Hsu}, D., \& {Haiman}, Z. 2020, arXiv
  e-prints, arXiv:2007.06529.
\newblock \doarXiv{2007.06529}

\end{thebibliography}
\bibliographystyle{aasjournal}

\end{document}